\documentstyle[aps,eqsecnum]{revtex}
\begin{document}
\title{Theory of semi-ballistic wave propagation} 
\author{A. Mosk, Th. M. Nieuwenhuizen}
\address{Van der Waals-Zeeman Instituut \\
Valckenierstraat 65-67, 1018 XE Amsterdam, The Netherlands\\}
\author{and C. Barnes}
\address{Cavendish Laboratory, University of Cambridge \\
Madingley road, Cambridge CB3 OHE, United Kingdom}
\date{\today}
\maketitle
\widetext
\begin{abstract}
Wave propagation through waveguides, quantum wires or films
with a modest amount of disorder is in the semi-ballistic regime  
 when in the transversal direction(s) almost no scattering occurs,
while in the long direction(s) there is so much scattering that
the transport is diffusive.

For such systems randomness is modelled by an inhomogeneous density
of point-like scatterers. These are first considered in
the second order Born approximation and then beyond that approximation.
In the latter case it is found
 that attractive point scatterers in a cavity always have
geometric resonances, even for Schr\"odinger wave scattering.   

In the long sample limit the transport equation is solved analytically.
Various geometries are considered:
waveguides, films, and tunneling geometries such as Fabry-P\'erot 
interferometers and  double barrier quantum wells.
The predictions are compared with new and existing 
numerical data and 
with experiment. The agreement is quite satisfactory.
\end{abstract}
pacs{\\ 
71.55 Jv (Disordered structures) \\
73.20 Dx (Electrons in low dimens. struct.)\\
73.50 -h (Electron transport in thin films)\\
42.81 Dp (Propagation in fiber optics)\\
73.40 Cg (Contact resistance)\\
73.40 G  (resonance tunneling)} 
%
\section{Introduction}
As is empirically known from the ancient development of music instruments,
cavity resonances determine the transmission of waves through devices 
which have dimensions of the order of the wavelength. 
The resonances give rise to transmission peaks of diverse systems such as
flutes, organ pipes,  Fabry-P\'erot interferometers, 
electronic  nanostructures \cite{BVHSSP}, and electronic waveguides.
In these systems the transmission of waves can drastically
increase if the wavevector of the incoming waves allows
for a new mode to be resonant. 
In the case of a very pure or very small cavity,
impurity scattering can be neglected and 
the transmission is said to be ballistic.
Ballistic transport has been shown to occur in various systems,
including quantum point contacts \cite{vHB} and narrow
optical slits \cite{MvdM}.

For wave propagation through waveguides or quantum wires
with a modest amount of disorder, several regimes occur.
For rather clean systems one still has ballistic transport of essentially
 unscattered waves. In the limit of
dirty, but non-absorptive and phase coherent systems, the intensity
mainly diffuses through the system. As pointed out recently by one of us,
for long wires or thin films
there is a third regime, the semi-ballistic regime. ~\cite{Theo}
Here disorder is large enough to cause 
diffusion in the long direction(s), but small enough
to maintain ballistic motion in the narrow directions.
In the present work we shall focus on this regime.

For electronic systems the conductance can be
 expressed by the Landauer formula,
\[G= \frac{2e^2}{h} \sum_{{ \bf a}{ \bf b}}{\cal T}^{\rm flux}_{{ \bf a}{ \bf b}}\]
in terms of the flux transmission coefficients
 ${\cal T}^{\rm flux}_{{ \bf a}{ \bf b}}$ 
of the system, where ${ \bf a}$ and ${ \bf b}$ stand for the incoming and outgoing
channels, respectively. 
In the ballistic regime there is no channel-to-channel scattering and
the transmission coefficients are diagonal in mode space.
They are of order unity for the propagating modes
(the low order cavity resonances that can be excited at
the energy of the
incoming wave), and exponentially small
for the evanescent modes (the higher order cavity resonances).
Whenever a new cavity mode becomes resonant, the conductance
makes a step of universal heighth $2 e^2/h$.

It is well known that multiple scattering by a large density
of impurities changes this ballistic transport to a diffusive one.
In the diffusive regime all
intensity is completely randomly distributed over the system.

In the intermediate regime of 
{\it semi-ballistic transport} a moderate amount of scatterers is present.
On one hand, the cavity modes of the pure system are hardly perturbed
while on the other hand
multiple scattering dominates transport in the long direction(s).
 
In such systems, interesting effects appear especially near
the onset of new resonances. The conductance of a quantum wire
shows a dip just before a new mode becomes resonant. \cite{MASEK}
\cite{ROTH} \cite{CHUSORB}. Certain GaAs/AlGaAs double barrier
quantum wells show diffusive broadening of their transmission resonance
\cite{GRMM}.

We study the average transmission properties of semi-ballistic
systems using the scalar wave approximation in
the limit of point-scatterers. 
We assume that neither finite-temperature effects nor Anderson 
localization are relevant.

In ref.\cite{Theo} one of the authors discussed a model to explain
the transport properties
of these systems. In the present work we will present the
derivations of the results
in that article and extend the approach 
beyond the second order Born approximation
to the t-matrix.

The setup of this paper is as follows:
In section \ref{wguidemod} we will calculate the transmission
coefficients of a long, moderately  disordered waveguide, to 
supply the derivation of the results presented in \cite{Theo}.
We then apply them 
to the calculation of the conductivity of disordered quantum wires.

In section \ref{ch-doublebarr} the transmission of semi-ballistic
double barrier structures is examined. 
Ballistic double-barrier structures have a transmission peaked at certain
discrete wavevectors. In the semi-ballistic regime there is a broadening
of these transmission resonances, and transmission to all 
other resonant channels.

In section \ref{ch-beyondborn} we  extend the
discussion to include resonance effects of the scatterers that may be
induced by the geometry.

In section \ref{ch-sim} we discuss the comparison between our model
and numerical simulations of an Anderson model in the semi-ballistic regime.
We close with a summary.

\section{Semi-ballistic transport in a waveguide}
\label{wguidemod}

We first consider the case of a moderately disordered
rectangular waveguide. To simplify the problem we consider scalar waves
instead of vector waves, thus neglecting polarization effects.
In this way we model the propagation of TE modes in electromagnetic waveguides.
The same approach also applies to electron propagation in quantum wires 
and, to some extent, sound propagation in long corridors. 

Our geometry has also been chosen as simple as possible. It consists
of an infinitely long waveguide with in the middle a section of
finite length in which a moderate amount of
scatterers is present (see fig. \ref{fig-wguide}).
Outside the disordered region we
have a clean system (perfect leads, for electronic systems)
in which ballistic transport occurs.
In the disordered region semi-ballistic transport occurs.
Here we consider (longitudinal) transport inside such a geometry, so that
the wall potentials can be considered infinitely high. 
In later sections we consider (tranverse) tunneling through
such a device and its boundaries. For that apllication the wall potentials
must have a finite strength.

\subsection{The pure system}
Our waveguide consists of four conducting plates, at $x=0$, $x=d_x$,
$y=0$ and $y=d_x$, and it is infinite in the $z$-direction.
The conducting plates impose the
 boundary condition $\Psi=0$ at their surface, to
describe TE waves:
\begin{eqnarray}
\label{plates}
\psi(0,y,z)=\psi(d_x,y,z)=0
\\
\psi(x,0,z)=\psi(x,d_y,z)=0 \nonumber
\end{eqnarray}
For electron propagation this would correspond to infinite wall potentials.
We assume that monochromatic waves of frequency
$\omega_0$, and free space wavevector $k_0$ propagate through the
guide. In the scalar wave approximation this is 
described by the following equation:
\begin{eqnarray}
\{-\nabla^2-k_0^2 \}\psi({\bf r})=0.
\end{eqnarray}
In the absence of impurities the waves travelling through the
guide have the form 
\begin{equation} 
\label{eigenf-1}
\psi(x,y,z)=
\sum_{{ \bf p}}\Psi_{{ \bf p}}(\bbox{\rho}){\rm e}^{\textstyle {\rm i} q_{{ \bf p}}z}
\end{equation}  
where $\bbox{\rho}=(x,y)$ is the transversal position and
$\Psi_{{ \bf p}}(\bbox{\rho})$ are the discrete transversal eigenmodes
 of the system:
\begin{eqnarray}
\label{modes}
\Psi_{{ \bf p}}(\bbox{\rho})&=&\sqrt{\frac{2}{d_x}}\sin({p_x} x)
\sqrt{\frac{2}{d_y}}\sin({p_y} y)
\\ \nonumber
p_x &=&\frac{m_x\pi}{d_x} \quad
p_y = \frac{m_y\pi}{d_y} \qquad
\left( m_{x,y}\ge1 \,; m_{x,y}\,\mbox{integer}\right)
\end{eqnarray}
Other geometries e.g., cylindrical waveguides, can be described in the same
way by substituting the corresponding mode wavefunctions for $\Psi$.
The modes for which ${ \bf p}^2<k_0^2$, are `free', i.e., they can 
propagate in the $z$-direction, with the mode-dependent 
wavenumber $q_{{ \bf p}}=\sqrt{k_0^2-{ \bf p}^2}$. 
The Green's function of this pure system, to be denoted
as $G^0$, is the solution of the equation
\begin{eqnarray}
\label{puregreen}
\{-\nabla^2-k_0^2 \}G^0({\bf r},{\bf r'})&=&\delta({\bf r}-{\bf r'})
\end{eqnarray}
with the boundary condidtions (\ref{plates}).
To incorporate the boundary conditions in the equation we
study the projections $G^0_{\bf p}$
\begin{eqnarray}
G^0_{{ \bf p}}(z,z')&=&\int {\rm d} \bbox{\rho}\; {\rm d} \bbox{\rho}'\;
\Psi^*_{{ \bf p}}({\bbox{\rho}})
G^0({\bbox{\rho}};z,{\bbox{\rho}'};z')
\Psi_{{ \bf p}}({\bbox{\rho}}')
\end{eqnarray}
Since we have translational invariance we can use a
Fourier transform to find from
(\ref{puregreen}):
\begin{eqnarray}
\label{fourierg0}
G^0_{\bf p}(q)=\frac{1}{q^2+{\bf p}^2-k_0^2-{\rm i} 0}
\end{eqnarray}
The extra term ${\rm i} 0$ ensures convergence of the back-transformation.
In real space one has:
\begin{eqnarray}
\label{Realsp0}
G^0_{{ \bf p}}(z-z') = 
\frac{{\rm i} \exp\{{\rm i} \sqrt{k_0^2-{{ \bf p}}^2}\,|z-z'|\}}
{2\sqrt{k_0^2-{{ \bf p}}^2}}
\end{eqnarray}

\subsection{Adding impurities}
Now suppose a small density of scattering impurities is present
at random positions ${ \bf R}_i$ in the
region \mbox{$0 \le z \le d_z$}. 
Their scattering properties can be expressed
in a scattering potential $V({ \bf r})$$=$
$\sum_i V_{\rm s}({ \bf r} -{ \bf R}_i)$.
We replace the single scatterer potential $V_{\rm s}$
by a $\delta$-potential with  scattering strength $-u$. This is known
to be a good approximation for electron scattering off a screened charge.
The density of scatterers 
\begin{equation}
\begin{array}{rcll}
 n({ \bf r})&=&n(x,y)\qquad &0<z<d_z\nonumber\\
              &=&0\qquad &z<0;\quad z>d_z 
\end{array}
\end{equation}
 needs not be homogeneous in the $x,y$ directions, we
will assume though that it is independent of $z$ for $0 \le z \le d_z$.
The density of scatterers will be assumed to be so small that
the cavity modes (\ref{modes}) are still well defined, i.e. the
scattering mean free path $\ell_{ \bf p}$ must be much larger than the 
transversal sizes $d_x,d_y$.
The wave equation for this system is
\begin{eqnarray}
\{-\nabla^2-u\sum_i\delta({{ \bf r}}-{\bf R}_i)\}\psi({{ \bf r}})&=&
k_0^2 \psi({ \bf r})
\end{eqnarray}
Here ${\bf R}_i$ are the random positions of the scatterers, distributed
according to a density $n({ \bf r})$. 

\subsection{The $t$-matrix of a single scatterer}
\label{ssec-tmatrsingle-1}

We first consider the case when only one scatterer is present
at position ${ \bf r}$.
The $t$-matrix of the point scatterer is simply the sum of the
Born-series expressing repeated 
scattering events at the same scatterer: 
\begin{eqnarray}
\label{t-expansie}
t({ \bf r}) &=& u+uG({ \bf r},{ \bf r})u+uG({ \bf r},{ \bf r})uG({ \bf r},{ \bf r})u+... \\ 
\label{fullt}
 &=&\frac{u}{1-uG({{ \bf r},{ \bf r}})}
\end{eqnarray}
Note that since the scatterer is pointlike, 
there is no momentum dependence and
the $t$-matrix is diagonal in real space.
Eq. (\ref{fullt}) depends on the return Green's 
function $G$, which
we have not expressed yet.
As $G$ is a property of the (local) environment
of the scatterer it is clear what the physical significance of
the $t$-matrix is: {\em the t-matrix describes
the effect of a scatterer in its local environment.}\ 
In the case where many randomly positioned scatterers are present,
the return Green's   function $G({ \bf r},{ \bf r})$ in $t$ depends on $t$ 
itself, which makes eq. (\ref{fullt}) self-consistent. 
 This means the $t$-matrix is not a property that can be taken from
 literature.  It must be calculated explicitly using the appropiate
return Green's  function in the system under consideration.
 
In one dimension $(d=1)$, equation (\ref{fullt}) is well defined.
For  $d \geq 2$ the real part of the return
Green's function diverges and a reinterpretation is needed.
Indeed, for a pure system in three dimensions the divergency is well
known from the equivalent of Coulomb's law
\begin{equation}
G({ \bf r},{ \bf r}')=\frac{{\rm e}^{ {\rm i}k|{ \bf r}-{ \bf r}'|}}
{4\pi|{ \bf r}-{ \bf r}'|}
\approx \frac{1}{4\pi|{ \bf r}-{ \bf r}'|} + \frac{{\rm i}k}{4\pi}
\end{equation}
As this divergent term appears in the denominator of (\ref{fullt}),
 the $t$-matrix vanishes, strictly speaking. 
This problem was discussed for scatterers in free space in
ref. \cite{NLT}.
We will examine this problem in more detail for constricted geometries
in section
\ref{ch-beyondborn}. 

For now we will proceed using the simplest way around this problem,
known as the second order Born approximation. This approximation is commonly 
used in electronic systems. Indeed, approaches with random potentials
 that obey Gaussian statistics are equivalent to the second order
Born approxiamtion. One of our aims is to see whether this is still a
good aprroximation in cavities. In this approach one has 
\begin{eqnarray}
\label{tborn}
t_{\rm Born} &=& u+{\rm i} u^2 {\rm Im \,} G({{ \bf r},{ \bf r}})
\approx {\rm i} u^2 {\rm Im \,} G({{ \bf r},{ \bf r}})\end{eqnarray}
The real part $u$ gives rise to a small
average potential and will be neglected from here on.
This approximation maintains the property of scattering but it 
does not take into account possible resonant behaviour
of the scatterer. 
In general, it is a good approximation for very weak scatterers.
In section \ref{ch-beyondborn} we will turn 
to the problem of including the full $t$-matrix in our calculations.
This will give rise to interesting resonance effects near the subband edge.

\subsection{The amplitude Green's  function and the selfenergy}
The self-energy 
$\Sigma({{ \bf r},{ \bf r}'})$ is defined as the sum of all irreducible 
scattering events that may be inserted in a Green's  function line. 
As the density of impurities is low, we can restrict ourselves to
the lowest order approximation to the average
self-energy, which is diagonal in the space coordinates,
 $\Sigma({ \bf r},{ \bf r}') = \delta({ \bf r}-{ \bf r}')
 \Sigma({ \bf r})$ with
\begin{eqnarray}
\Sigma({{ \bf  r}})&\approx&n({\bbox{\rho}})t_{\rm Born}({{ \bf  r}})=
iu^2n(\bbox{\rho}){\rm Im}G({ \bf r},{ \bf r})
\end{eqnarray}
The average Green's function G is expressed in the 
(average) self-energy by the Dyson equation:
\begin{equation}
\label{dyson}
G({{ \bf  r}},{\bf {r}}')
=
G^0({{ \bf r}},{{ \bf r}}')
+\int {\rm d}^3{{ \bf  r}}'' G^0({{ \bf  r}},{{ \bf  r}}'')
\Sigma({{ \bf r}}'')G({{ \bf  r}}'',{{ \bf  r}}')
\end{equation}
To deal with the multiple scattering problem we have to average
over impurity positions, weighted with their density $n(\bbox{\rho})$. 
Within  the second order Born 
 approximation we find an explicit form for the 
 Green's functions $G_{{ \bf p}}$ of the mode ${ \bf p}$
\begin{eqnarray}
\label{greenp}
G_{{ \bf p}}(q)&=&\frac{1}{{{ \bf p}}^2+q^2-k_0^2- \Sigma_{\bf p}}
\end{eqnarray} 
In the second order Born approximation
$\Sigma_{\bf p}$$=$${\rm i}\Gamma_{\bf p}$, so the resonance width reads
\begin{eqnarray} 
\Gamma{{ \bf p}}&=&
\int {\rm d}^2 \bbox{\rho}\, n(\bbox{\rho})\,{\rm Im}\;t_{\rm Born}(\bbox{\rho})
\Psi_{\bf p}^2(\bbox{\rho})
\\ \nonumber
&=&\int {\rm d}^2 \bbox{\rho}\, u^2n(\bbox{\rho})\,{\rm Im \,}
G(\bbox{\rho},z,\bbox{\rho},z)\Psi_{{ \bf p}}^2(\bbox{\rho})
\\ \nonumber
&=&\int {\rm d}^2 \bbox{\rho}\, u^2n(\bbox{\rho})\,
\sum_{{ \bf p}'}{\rm Im\,}G_{{ \bf p}'}(z,z)
\Psi_{{ \bf p}'}^2(\bbox{\rho})\Psi_{{ \bf p}}^2(\bbox{\rho})
\end{eqnarray}
The result for $\Gamma_{{ \bf p}}$ does not depend on $z$, since
after averaging we have translational invariance when we are far away from
the leads at $z=0$ and $z=d_z$.
The form (\ref{greenp}) for $G$ yields in real space 
\begin{eqnarray}
\label{Realspace}
G_{{ \bf p}}(z,z') &=& \frac{{\rm i} \exp \{
{\rm i} \sqrt{k_0^2-{ \bf p}^2+i\Gamma_{{ \bf p}}}\,|z-z'|\} }
{2\sqrt{k_0^2-{{ \bf p}}^2+i\Gamma_{{ \bf p}}}}
\\ \nonumber
&=& \frac{{\rm i}\, {\rm e}^{\textstyle  {\rm i} q_{{ \bf p}}|z-z'|}}
{2 q_{{ \bf p}} + {\rm i}/\ell_{{ \bf p}}} \, {\rm e}^{\textstyle  - |z-z'|/2\ell_{{ \bf p}}}
\\ \label{def-l-and-q}
\mbox{with}&&q_{{ \bf p}}={\rm Re}\,\sqrt{k_0^2-{\bf p}^2+i\Gamma_{\bf p}}
\\ \rm{and} &&
 \ell_{{ \bf p}}=\frac{1}{2 {\rm Im} \sqrt{k_0^2-{\bf p}^2+i\Gamma_{\bf p}}}
\end{eqnarray}
The quantity $\ell_{{ \bf p}}$ is 
the mode-dependent elastic mean free path.
We find the selfconsistent equation
\begin{eqnarray}
\Gamma_{{ \bf p}}&=&\Gamma^{D}_{{ \bf p}}
\\ \label{Gamma}
\Gamma^{D}_{{ \bf p}}&\equiv&\sum_{{{ \bf p}}'}N_{{ \bf p}{ \bf p}'}
{\nu}_{{ \bf p}'}\label{GammaD=}
\\ 
\label{nu}
{\nu}_{{ \bf p}}(k_0)&\equiv&\,{\rm Re} \frac{1}
{2 \sqrt{k_0^2-{ \bf p}^2+{\rm i} \Gamma_{{ \bf p}}}}
=
\frac{\frac{1}{2} q_{{ \bf p}}}
{|k_0^2-{ \bf p}^2+{\rm i} \Gamma_{{ \bf p}}|}
\\
N_{{ \bf p}{ \bf p}'}&\equiv& u^2 \int {\rm d}^2 \bbox{\rho}\,\,n(\bbox{\rho})
\Psi_{{ \bf p}}^2(\bbox{\rho})\Psi_{{ \bf p}'}^2(\bbox{\rho})\label{Npp=}
\end{eqnarray}
The number of states (per unit length) 
in a mode ${\cal{N}}_{ \bf p}$ is 
\begin{equation}
\nonumber
{\cal{N}}_{ \bf p}=\frac{1}{\pi}{\rm Re}\; \sqrt{k_0^2-{ \bf p}^2+{\rm i}\Gamma_{ \bf p}}
\end{equation}
from which we can see that $\nu_{ \bf p}$ is proportional to the
density of states
\begin{equation}
\nu_{ \bf p}=\frac{\pi}{2k_0} \frac{{\rm d}{\cal{N}}_{ \bf p}}{{\rm d}k_0}
\end{equation}

\subsection{The Bethe-Salpeter equation}

To describe transport of intensity, electromagnetic energy or the 
probability  of Schr\"odingers particles 
through the system, we need the averaged intensity Green's function,
\begin{equation}
H({ \bf r},{ \bf r}')=\overline{G({ \bf r},{ \bf r}')G^*({ \bf r},{ \bf r}')} \end{equation}
In our model of discrete
eigenmodes we consider the projection of the intensity Green's
function $H_{{ \bf p}{ \bf p}'}$.
It describes the propagation of {\em intensity} from
 mode ${ \bf p}$ to ${{ \bf p}'}$, and it obeys the following
 Bethe-Salpeter (BS) equation:
\begin{eqnarray}
\label{twoparticle-h}
H_{{ \bf p}{ \bf p}'}(z,z') &=& 
G_{{ \bf p}}(z,z')G^*_{{ \bf p}}(z,z')\delta_{{ \bf p}{ \bf p}'}+ 
\\ \nonumber 
         && \sum_{{ \bf p}''} \int {\rm d} z''
           G_{{ \bf p}}(z',z'')G_{{ \bf p}}^*(z',z'') \times
\\ \nonumber
           U_{{ \bf p}{ \bf p}''}
           H_{{ \bf p}''{ \bf p}'}(z'',z')
\end{eqnarray}
This equation involves the irreducible vertex $U_{{ \bf p} { \bf p}'}$.
In our situation it is  independent of $z$ for $0<z<d_z$, while
it vanishes in the `leads' $-\infty<z<0$ and $d_z<z<\infty$. 
The irreducible vertex can be shown to be
the sum of all two-particle irreducible 
diagrams that can be inserted in the {\em intensity}
Green's  function. Two-particle irreducible in 
this context means that the diagrams cannot be
split into two separate diagrams by cutting one propagator and one complex
conjugated propagator line.

The irreducible vertex is not available in closed form so it must be
 approximated. We are however {\em not} free in choosing how to approximate
 it: the approximation must be consistent with the approximation to $\Sigma$
 we made earlier. This can be understood as follows: $U$ describes the emission
 of diffuse intensity by the scatterers, $\Sigma$ describes the intensity
 extinction due to the scattering. 
 If the two are not balanced, our description of the system
 will show {\em gain} or {\em absorbtion}, which is certainly unphysical in
 the systems we consider here.
 Flux conservation is guaranteed by the Ward-Takahashi identity which can
 be derived from field theory or by 
 manipulation of diagrams \cite{Mahan},\cite{Wolfle}.
\begin{eqnarray}
\label{Ward}
{\rm Im\;} \Sigma^{\rm D}_{{ \bf p}}
&=&
\sum_{{ \bf p}'}
U_{{ \bf p}{ \bf p}'}
{\rm Im \;} G_{\bf p'}(z,z)
\end{eqnarray}
If this identity holds, flux is conserved to every order 
in the scatterer density
while most of our other approximations are valid only 
in leading order in density.
Comparing (\ref{Ward}) to (\ref{Gamma}) shows that
\begin{eqnarray}
\label{vertex}
U_{{ \bf p} { \bf p}'}&=&N_{{ \bf p} { \bf p}'}
\end{eqnarray}
satisfies the Ward-Takahashi identity. It is, in fact, the
ladder vertex, constituting one step in a ladder diagram. 
It  is known from transport theory that the ladder diagrams
describe diffusive transport.

\subsection{Solving the transport equation in a waveguide}

Since we cannot solve the Bethe-Salpeter equation analytically
we will try to gain as much information as possible from
approximations. The quantities we are interested in are the average
longitudinal intensity transmission coefficients $T_{{ \bf a}{ \bf b}}$,
 where we
use ${ \bf a}$ for the transverse momentum of an incoming cavity mode
and ${ \bf b}$ for an outgoing mode.

We consider a  wave coming in from $z=-\infty$ of the
form $\psi_{\rm in}({ \bf r})=\Psi_{{{ \bf a}}}(\bbox{\rho})
\exp({\rm i} q_{{{ \bf a}}}z)$. 
It will be attenuated upon entering the disordered region. This is described 
by the amplitude Green's  function. In the disordered section it 
gives rise to a source intensity
\begin{equation}
\label{s-term}
S_{{ \bf p}}(z)=\delta_{{{ \bf p}},{{ \bf a}}}{\rm e}^{\textstyle -z/\ell_{{ \bf p}}} 
\end{equation}  
We have neglected possible surface reflections as we assume the dispersion
relation to be the same inside the disordered region as outside.
The total intensity present in any mode as a function of position
is denoted $\Phi_{{ \bf p}}(z)$. Using (\ref{twoparticle-h}) it can be shown
that this quantity obeys the ladder equation
\begin{eqnarray}
\label{KaleLadder}
\Phi_{{ \bf p}}(z)=S_{{ \bf p}}
+  \int_0^{d_z} {\rm d}z'G_{{ \bf p}}
(z,z')G^*_{{ \bf p}}(z,z') \times
\\ \nonumber
\sum_{{ \bf p'}} U_{{ \bf p}{ \bf p'}} \Phi_{{ \bf p}'}(z')
\end{eqnarray}
We can re-express this by expanding the Green's functions
\begin{eqnarray}
\label{phi-eq}
\Phi_{{ \bf p}}(z)=   S_{{ \bf p}}(z) + 
\\ \nonumber
\frac{ \nu_{{ \bf p}}} {2 \Gamma_{{ \bf p}} \ell_{{ \bf p}}}
\int_0^{d_z} {\rm d} z' {\rm e}^{\textstyle -|z-z'|/\ell_{{ \bf p}}}
\sum_{{ \bf p}'} U_{{ \bf p}{ \bf p}'}\Phi_{{ \bf p'}}(z')
\end{eqnarray}
where we have inserted the relation
\[
\frac{1}{|k_0^2-{{ \bf p}^2}+{\rm i}\Gamma_{{ \bf p}}|}
=
\frac{2 \nu_{ \bf p}}{\Gamma_{{ \bf p}} \ell_{{ \bf p}}}
\]
Equation (\ref{phi-eq}) is a linear system of
Fredholm integral equations of the second kind.
The solution of this type of equation, for $d_z \to \infty$,
 is the sum of a homogeneous
solution, $\Phi^{\rm H}$,
and a special solution $\Phi^{{{ \bf a}}}$ which is dependent on the
source term $S_{{{ \bf a}}}$.
 (In fact, the special solution is defined except for a multiple
of the homogeneous solution which we can add to it. We will choose
$\Phi^{{{ \bf a}}}$ such that
it remains finite as $z \to \infty$.)

By differentiating the system of Fredholm equations we find
the following equivalent set of differential equations:
\begin{eqnarray}
\label{diff-eq}
\Phi_{{ \bf p}}''(z)&=& \frac{1}{\ell_{{ \bf p}}^2}\Phi_{{ \bf p}}(z)
- \frac{\nu_{{ \bf p}}}{\Gamma_{{ \bf p}}\ell^2_{{ \bf p}}}
\sum_{{ \bf p}'} 
U_{{ \bf p}{ \bf p}'} 
 \Phi_{{ \bf p}'}(z) 
\\
\label{bound-0}
\ell_{{ \bf p}}\Phi_{{ \bf p}}'(0)
&=& - 2 S_{{ \bf p}}(0) - \Phi_{{ \bf p}}(0)
\\
\label{bound-d3}
\ell_{{ \bf p}}\Phi_{{ \bf p}}'(d_z)
&=&  \Phi_{{ \bf p}}(d_z)
\end{eqnarray}

We will first study the solution to the homogeneous form of the system
(\ref{diff-eq}), that is to say, we take $S=0$ and $d_z\to \infty$. Then we
have only the boundary condition (\ref{bound-0}) at $z=0$ .
This will then be used to construct to solution for finite $d_z$.

Using eq. (\ref{Gamma}) 
 the Ward identity (\ref{Ward}) can be written as 
\begin{equation} 
\sum_{{ \bf p}'}U_{{ \bf p}{ \bf p}'}\nu_{{ \bf p}'}=\Gamma_{{ \bf p}}
\end{equation}
This implies that the right hand side of eq. (\ref{diff-eq}) vanishes 
if we insert $\Phi_{{ \bf p}}(z)\propto \nu_{{ \bf p}}$.
The differential equation therefore has a solution of the form
\begin{equation}
\label{phi-homogeen}
 \Phi^{\rm H}_{{ \bf p}}(z)=(z_0+z)\nu_{{ \bf p}} ,\,\,\,{\rm for}\, z>>\ell_{ \bf p}
 \end{equation}
 Near the boundary there will be other terms because of the condition
 (\ref{bound-0}). They are related to the non-zero eigenvalues
 of the matrix, so they decay exponentially away from the edges.
 The asymptotic behaviour of the homogeneous solution
 is characteristic of one-dimensional diffusion:
 the intensity decreases linearly with $z$. 
 As expressed by the factor $\nu_{{ \bf p}}$ in (\ref{phi-homogeen}),
 the intensity is distributed over the modes according to their
 density of states.
 The shift $z_0$ will be calculated further on, when we take in to account the
boundaries.

  The special solution for an incoming wave of unit intensity,
  $\psi^{\rm in}$$=$${\rm e}^{\textstyle {\rm i}q_z z}\Psi_{{ \bf a}}$ in mode
 ${ \bf a}$ is called $\Phi^{{ \bf a}}$. In the case of a semi-infinite system
 we can choose it such, that it converges to a constant  
 away from the boundary. The distribution over the modes is then
 given by
 \begin{equation}
  \label{def-special}
 \Phi^{{ \bf a}}_{{ \bf p}}(z) \to C_{{ \bf a}}\nu_{{ \bf p}} \, ,\, z>>\ell_{{ \bf p}}
 \end{equation}  
 The coefficient $C_{{ \bf a}}$ is different for each incoming mode.

We now examine the behavior of the solutions to (\ref{diff-eq}) in terms of
the eigenvalues of the matrix of the system.
There are exponentially growing solutions, exponentially decaying ones and
linear+constant solutions corresponding to the zero eigenvalue of the
system.
For a semi-infinite system, the exponentially growing 
solutions will be absent. 
The equation can then be solved formally, yielding
\begin{equation} 
\Phi_{\bf p}(z)=\sum_{i} c_{i} R^{i}_{\bf p} {\rm e}^{\textstyle -z \lambda_{i}}
+(\alpha+\beta z)R^{0}_{\bf p}
\end{equation} 
where $R^{i}_{ \bf p}$ are the right eigenvectors of the system
(\ref{phi-homogeen})
and all eigenvalues $\lambda_i$ are positive.
The linear plus constant term corresponds to the eigenvalue zero
of the system, with the right-eigenvector  $R^0_{\bf p}=\nu_{\bf p}$.

The boundary condition at $z=0$ puts constraints on the coefficients
$\alpha$,$\beta$ and $c_i$. For the homogeneous solution defined in 
(\ref{phi-homogeen}) the boundary condition is
\begin{equation} 
\ell_{\bf p} \Phi_{\bf p}'(0) = \Phi_{\bf p}(0)
\end{equation} 
according to this definition, $\beta=1$ and $\alpha=z_0$
which leads to the equation for the coefficients of the homogeneous
solution $c^{\rm H}_i$:
\begin{equation}
\label{eq-ch}
\sum_i c^{\rm H}_i r^i_{\bf p} (\lambda_i\ell_{\bf p} +1 ) +z_0 R^0_{\bf p}
=R^0_{\bf p} \ell_{\bf p}
\end{equation} 

For the special solution to the problem of a source intensity in channel
${ \bf a}$ the definition (\ref{def-special}) 
leads to $\alpha=C_{{ \bf a}}$ and $\beta=0$.
The resulting equation for the coefficients $c^{{ \bf a}}_i$ of the
special solution reads

\begin{equation}
\label{eq-ca}
\sum_i c^{{ \bf a}}_i R^i_{\bf p} (\lambda_i\ell_{\bf p} +1 ) +C_{{ \bf a}} R^0_{\bf p}
=2\delta_{{ \bf a},{\bf p}}
\end{equation} 

This equation is very similar to (\ref{eq-ch}), if we take the sum
$\sum_{{ \bf a}} R^0_{{ \bf a}}\ell_{{ \bf a}} ...$ on both sides of
(\ref{eq-ca}) the equations become identical and we can conclude
\begin{equation}  
\label{eq-z0}
z_0=\frac{1}{2} \sum_{{ \bf a}} \nu_{{ \bf a}} \ell_{{ \bf a}} C_{{ \bf a}}
\end{equation} 

A very useful sum rule can be found from this equation by
multiplying (\ref{eq-ca}) by $R_{{ \bf a}}^0$ on both sides and summing:
\begin{equation}
\sum_{i,{ \bf a}} R^0_{{ \bf a}} c^{{ \bf a}}_i R^i_{\bf p} (\lambda_i\ell_{\bf p} +1 )
+ R^0_{{ \bf a}} C_{{ \bf a}} R^0_{\bf p}
=2 R^0_{{ \bf p}}
\end{equation} 
Taking the inproduct with a vector orthogonal to the leftmost
term, but not to $R^0_{ \bf p}$, we find:
\begin{equation}
\label{sumrule-1} 
\sum_{{ \bf a}} \nu_{{ \bf a}} C_{{ \bf a}} = 2
\end{equation} 
To find another useful constant we study the two-particle Green's function
$H$. From the time reversal symmetry of the problem we can derive an
  useful identity:
 \begin{equation}
 \label{timerev-1}
 {H}_{{ \bf p}{ \bf p}'}(z,z')={H}_{{ \bf p}'{ \bf p}}(z',z)
 \end{equation}
 If we let $z'$ become large, the Green's  function will behave like the
 homogeneous solution $\Phi^{\rm H}$ near
 $z=0$ because all other contributions are extinguished within 
 a few mean free paths
  from $z'$.
 From the symmetry property (\ref{timerev-1})
  and the behaviour at large $z$
  we find the mode distribution at $z'$ must be proportional
  to $\nu_{{ \bf p}}$:
 \begin{equation}
 \label{limiet-h-z}
 \lim_{z' \to \infty} H_{{ \bf p}{ \bf p}'}(z,z')=
 C_0 \nu_{{ \bf p}'} \Phi^{\rm H}_{{ \bf p}}(z)
 \end{equation}
To find
an expression for the coefficient $C_0$
 we consider a point source in mode ${ \bf a}$ at a large distance
$z_1 >> \ell$ from the boundary, so that all contributions 
that correspond to the nonzero eigenvalues of the system
 will have damped out there. It follows we only have to
consider the contribution that corresponds to the zero eigenvalue,
which yields, by considering the jump in 
the derivative, 
\begin{equation} 
\label{c0-regel}
C_0=\left\{\sum_{\bf p} \nu_{\bf p} \ell_{\bf p} \Gamma_{\bf p} \right\}^{-1}
\end{equation} 
Using this coefficient we find an expression for $C_{{ \bf p}}$:
\begin{equation}
C_{{ \bf p}}=\int_{0}^{\infty} {\rm d}z\; 
{\rm e}^{\textstyle -z/\ell_{{ \bf p}}} C_0
\sum_{{ \bf p}'} U_{{ \bf p}{ \bf p}'} \Phi^{\rm H}_{{ \bf p}'}(z)
\end{equation}
leading to the interesting relation
\begin{equation}
\label{phinull}
\Phi^{\rm H}_{{ \bf p}}(0)=
\frac{\nu_{{ \bf p}}}{2\Gamma_{{ \bf p}}\ell_{{ \bf p}}C^0} C_{{ \bf p}}
\end{equation}

\subsection{Transmission coefficients}
If the sample is finite a certain fraction
of the intensity that enters the sample at
$z=0$ will be transmitted to $z=d_z$.
We can calculate the transmission coefficients
from channel ${{ \bf a}}$ to channel ${{ \bf b}}$ for 
`optically thick' samples 
(length of many mean free paths) by
matching the solution of the ladder equation 
near both boundaries:
\begin{itemize}
\item For  $0 \ge z$ but $z-d_z >> \ell_{ \bf p}$
 the solution wil be the sum of the
special solution and a multiple of the
homogeneous solution
\begin{eqnarray}
\Phi_{{ \bf p}}(z) = -c\,\Phi^{\rm H}_{{ \bf p}}(z)
+ 
\\ \nonumber
\sum_{{ \bf p}'}
\int_0^\infty {\rm d}z'\,
  H_{{ \bf p}{ \bf p'}}(z,z')
  U_{{ \bf p}' {{ \bf a}}}
  {\rm e}^{\textstyle -z'/\ell_{{{ \bf a}}}}
\end{eqnarray}
\item For $z \le d_z$ but $z >> \ell_{{ \bf p}}$ the problem can be
 considered from $z=L$.
 There is no incoming 
intensity and only the homogeneous solution will be present,
\begin{equation}
 \Phi_{{ \bf p}}(z) = c\,\Phi^{\rm H}_{{ \bf p}}(d_z-z) 
\end{equation}
\end{itemize}
 In the bulk both solutions have a linear+constant form.
This makes it possible to match a (special+homogeneous)
solution at $z=0$ to a (homogeneous) solution at $z=d_z$.
\begin{eqnarray}
\label{fitting-edge}
\Phi^{\rm bulk}_{{ \bf p}}(z) &=&
 \left( C_{{{ \bf a}}} -c(z+z_0)\right) \nu_{{ \bf p}}
\\ 
\Phi^{\rm bulk}_{{ \bf p}}(z) &=& c(d_z-z+z_0) \nu_{{ \bf p}}
\end{eqnarray}
It follows that 
\begin{eqnarray}
c&=& \frac{1}{(d_z+2z_0)} C_{{{ \bf a}}}
\end{eqnarray}
Since the average Green's  function extinguishes in few mean
free paths, we do not have to take into account the
precise behaviour of the intensity at $z=0$ for
calculating the transmission to $z=d_z$. As usual, since
the sample's length is many mean free paths,
the transmitted fraction of the unscattered intensity
is negligible. The  intensity
transmission coefficient for transmission from channel $a$ to
channel $b$ is then equal to the intensity $\Phi_{{ \bf b}}(d_z)$,
\begin{eqnarray}
\label{Transition-matrix}
\nonumber
T_{{{ \bf a}}{{ \bf b}}}&=&\int_{0}^{d_z} {\rm d} z
G_{{{ \bf b}}}(z,d_z) G^*_{{{ \bf b}}}(z,d_z)
\sum_{{ \bf p}} U_{{ \bf b}{ \bf p}} \Phi_{{{ \bf p}}}(z)
\\ \nonumber
&\approx& \frac{C_{{{ \bf a}}}}
{(d_z+2z_0)}
\Phi^{\rm H}_{{ \bf b}}(0)
\\ \nonumber
&=& \frac{C_{{{ \bf a}}} C_{{{ \bf b}}}}
{4(d_z+2z_0)q_{{{ \bf b}}}^2}
\sum_{{ \bf p}} \Gamma_{{ \bf p}} \ell^2_{{ \bf p}} \nu_{{ \bf p}}
\end{eqnarray}
We have now derived formula (4) of ref. \cite{Theo}.
It holds for the transmission of scalar waves through waveguides and can
mutatis mutandis be applied to the propagation of EM waves or
Schr\"odinger waves. Below it is used to calculate the conductance of a
quantum wire.

\section{Conductance of electronic systems}

We now apply our result to the electronic case of a 
disordered conducting channel.
The Landauer formula gives the average zero temperature conductance
of a sample of arbitrary dimensions connected to two reservoirs
of electrons,
in terms of the average flux transmission coefficients:
\begin{eqnarray} 
\label{landauer}
R^{-1}=G&=&\frac{2e^2}{h}\sum_{a,b}{\cal T}^{\rm flux}_{ab}
=\frac{2e^2}{h}\sum_{a,b}\frac{q_b}{q_a} {T}_{ab}
\end{eqnarray} 
where $2 e^2/h$ is the quantum of conduction, the factor 2 comes
from spin degeneracy.
The flux transmission coefficients differ in our case from the
intensity transmission coefficients by a factor $q_{b;0}/q_{a;0}$, where
the $q_0$'s stand for the $z$-wavenumbers of the incoming and outgoing waves
outside the disordered region. To a good approximation it holds for the propagating
modes,
\begin{equation}
q_{{{ \bf a}};0}\approx\sqrt{q_{{ \bf a}}^2+\ell_{{ \bf a}}^{-2}}
\approx \frac{1}{2 \nu_{{{ \bf a}}}}
\end{equation} 
With these approximations and relation (\ref{sumrule-1}), we find
\begin{eqnarray} 
G&=&\frac{2e^2}{h}
\frac{4}
{(d_z+2z_0)} 
\sum_{{ \bf p}} \Gamma_{{ \bf p}} \ell^2_{{ \bf p}} \nu_{{ \bf p}}
\end{eqnarray} 

>From this formula, it is easy to see that Ohm's law is valid
for the average conductance of samples in the semi-ballistic regime.
The resistance, defined as the inverse of the average conductance,
reads~\cite{NLUCK}
\begin{equation}
\label{rplusrc}
R=\frac{d_z}{d_xd_y} \frac{1}{\sigma} + 2 R_{\rm c}
\end{equation} 
with the conductivity
\begin{equation} 
\label{sigmabulk}
\sigma=\frac{8e^2}{hd_xd_y}\sum_{{ \bf p}}
\Gamma_{{ \bf p}} \ell_{{ \bf p}}^2 \nu_{{ \bf p}}
\end{equation}
and the `contact resistances' at $z=0$ and $z=d_z$
\begin{equation} 
R_{\rm c}=\frac{z_0}{\sigma}
\end{equation} 
Note, however, that $R_{\rm c}$ as well as $\sigma$ are 
complicated functions of $k_0$.

In ref. \cite{Theo}
an analytic result was obtained for the conductance in a $2D$ film,
with the width $d_y \to \infty$, in the limit of weak disorder:
\begin{equation}
\frac{\sigma(d_x)}{\sigma_{\rm bulk}}=\frac{3 N}{2N+1}
-\frac{1}{2}N(N+1)\frac{(d^*)^2}{d_x^2}
\end{equation}
where $d^*=\pi/k_0$ is the resonant width, $N$ is the number of
open channels and $\sigma_{\rm bulk}=\frac{2e^2}{3\pi h} k_f^2 \ell_{\rm bulk}$
is the Drude conductivity of a 3D bulk sample.
This result is reproduced here in figure
(\ref{Theofig}), together with the result for a $1D$ quantum wire.
Both curves have been scaled with the bulk
conductivity.

It is seen that these curves exhibit remarkable drops in the conductivity
whenever a new  cavity mode becomes resonant. 
These drops are explained mathematically by the density of states
that grows large near the subband bottom, thus causing the second order
Born $t$-matrix to become larger. This expresses more efficient
scattering and, therefore, less conductance.
The physical explanation is: When the `new' mode is not yet resonant
it does not yet contribute to conduction. There is scattering to this
mode however, and the scattered waves interfere destructively with the
waves present in the other modes.
In section \ref{ch-beyondborn} we will study the analogue of this effect
for the full $t$-matrix of the pointscatterers.

A different approach for calculating the conductivity was followed by
Surke and Wilke in ref \cite{Surke}. These authors calculate the
conductivity directly from the Kubo formula and derive expressions different
from ours. Their results involve the average scattering time, rather than the
inverse of the average scattering rate ($\Gamma_{{ \bf p}}$ in our work).
It therefore seems to us, that the latter results are unphysical.
Indeed, one can consider cases where the scatterer density goes to zero locally,
such that the average scattering time diverges.
Then the resulting prediction for the conductivity diverges, while our result
(\ref{sigmabulk}) is quite insensitive to such limits, as it should be.


\section{Transport through a double barrier structure}
\label{ch-doublebarr}

In section \ref{wguidemod}, transmission coefficients were derived for
 transport of wave intensity
along the length of a waveguide.
 In some systems, like
the Fabry-Perot Interferometer (FPI), or its electronical analogue,
the double barrier quantum well (DBQW), transport occurs in the
transversal direction due to `tunneling' through the barriers.

In the absence of random scattering these devices transmit only waves
 for which the perpendicular component of the 
wavevector is resonant with the
cavity. The linewidth is very small, usually the $Q$-factor
of these devices is several thousands. 
The pure FPI transmits a light beam that meets the resonance condition
without changing its direction. 

The devices we are interested in contain a small density of impurities,
such that the width $d_x$ of the device is still much smaller than
the mean free path, but the width multiplied by the $Q$ factor
is much larger than the mean free path.
The problem of multiple scattering in such devices was first
considered on a fundamental level by Berkovits and Feng \cite{BF}.
 Their approach is valid for the situattion of one resonant mode, 
well away from the onset of further cavity resonances.

The behaviour near resonances in the multimode situation
was later discussed by one of us \cite{Theo}.
In the present section we will discuss
the derivation of these results.  
It will be seen that multiple elastic scattering will broaden the
resonance linewidth and cause transmission of the energy into 
all available outgoing
channels, independent of the incoming channel.

\subsection{Double barrier system in one dimension}

First, for simplicity, we will consider a one-dimensional  
double barrier quantum well, like in fig. (\ref{fig-dbqw}),
with the potential between the barriers equal to that outside.
We will describe these barriers, which are imperfect 
mirrors, by  strong $\delta$-function potentials.
Their strengths are allowed to have an imaginary part  to
model absorbtion of waves by the mirrors.

\begin{equation} 
V(x)=V_{\rm L}\delta(x)+V_{\rm R} \delta(x-d_x)
\end{equation} 

The intensity transmission, reflection and absorbtion coefficients
of the single barriers follow from requiring continuity of the wavefunction
across the barrier:
\begin{equation}
T=
\frac{4k_0^2}{|V+2{\rm i}k_0|^2}
\qquad
R=\frac{|V|^2}{|V+2{\rm i}k_0|^2}
\end{equation}
If $V$ is purely real, there is no absorbtion and the
identity $T+R=1$ holds. If ${\rm Im}\; V > 0$ the model describes a
barrier with nonzero absorbtion coefficient $A$:
\begin{equation} 
A=1-R-T=\frac{4k_0{\rm Im}\; V }{|V+2{\rm i}k_0|^2}
\end{equation} 
A description with absorption is realistic in the optical case,
 where the absorbtion coefficient of a mirror
may be larger than the transmission coefficient.
In the electronic case there should be no absorbtion, but there could be 
loss of phase coherence which can lead to similar effects.
 
Now we look at the Green's function inside the structure,
$G(x,x')$ for $0<x,x'<d_x$.
By requiring continuity at both barriers we can see it is,
neglecting terms of order $1/V^2$,
\begin{equation}
\label{fpisin}
G(x,x')=
- \frac{2{\rm i}}{k_0}
\frac{
 \sin(k_0 x +\delta) \sin(k_0 x'+\delta)
}
{ \sin (k_0 d_x) 
+\frac{2{\rm i}k_0}{V_L-2{\rm i}k_0}
+\frac{2{\rm i}k_0}{V_R-2{\rm i}k_0}
}
\end{equation}
where $\delta$ is a (complex) phase shift which becomes small near resonances.
In the ideal system ($V$ real and very large) these resonances
 occur when $k_0={n\pi}/{d_x}$.
If the linewidth of the structure is much smaller than the mode separation,
which is the case for high-Q cavities, we can make a Lorentzian approximation
to the lineshape and express the propagator as a sum over mode wavefunctions:
\begin{equation}
G(x,x')=\sum_{\textstyle p_x=\frac{n\pi}{d_x}} 
\Psi_{p_x}(x) G_{p_x}(k_0) \Psi_{p_x}(x')
\end{equation}
involving the mode wavefunctions
\begin{equation}
\label{psi1ddb}
\Psi_{p_x}(x)=\sqrt{\frac{2}{d_x}}\sin(p_x x + \delta_{p_x})
\end{equation}
and the mode-propagator, (upon neglecting small shifts of the resonance positions)
\begin{equation}
\label{g1ddb}
G_{p_x}(k_0)=\frac{1}{p_x^2-k_0^2 -{\rm i} \Gamma^{\rm pure}_{p_x}}
\end{equation}
The linewidth $\Gamma^{\rm pure}$ is due to absorbtion and transmission
of the barriers:
\begin{equation}
\Gamma^{\rm pure}_{p_x} =
\frac{p_x}{2d_x}(  A_{L}+ T_{L}+ A_{R}+T_{R})
\end{equation}
where $L,R$ stands for left and right, respectively.
This expression of the linewidth in terms of  transmission and
absorbtion coefficients is valid for general barrier potentials.

A  beam coming in from $x=-\infty$ of the form $\exp({\rm i}k_0x)$,
with $k_0$ close to some resonant $p_x$
will give rise to a standing wave inside the structure. The amplitude
of this wave is found by matching at the boundaries, which gives,
approximating for $k_0$ near $p_x$ :
\begin{eqnarray} 
\psi(x) &=& \frac{-2{\rm i}k_0^2}{V_L-2{\rm i}k_0}
 \Psi_{p_x}(x)G_{p_x}\sqrt{\frac{2}{d_x}}
\end{eqnarray}

To describe realistic situations we need to  expand the simple 
$1D$ model to a three dimensional model 
with barriers (`mirrors') at $x=0,x=d_x$.
There are three regimes for this generalization:
\begin{itemize}
\item A structure which is confined in both y and z dimensions, like
a double barrier structure inside a waveguide.
\item A structure which is confined in the y direction but very large in
z dimension, for example a double barrier structure inside a 2D electron gas.
\item A structure which is very large in both y and z dimensions, like
an optical FPI or a double barrier quantum well inside
a three dimensional semiconductor sample.
\end{itemize}
The first case is the simplest to describe,
we will analyze it first and then take
the continuum limit in two directions.
Results for the pure case follow from simply replacing the
1D mode eigenfunctions with the 3D eigenfunctions and replacing
the mode energy $p_x^2$ by ${ \bf p}^2=p_x^2+p_y^2+p_x^2$
\begin{eqnarray}
G({ \bf r},{ \bf r'}) &=&
\sum_{{ \bf p}=p_x,p_y,p_z}
\Psi_{{ \bf p}}({ \bf r})
G_{{ \bf p}}(k_0)
\Psi_{{ \bf p}}({ \bf r}')
\\
\Psi_{{ \bf p}}({ \bf r})
&=&\sqrt{\frac{8}{d_xd_yd_z}}\sin(xp_x)\sin(yp_y)\sin(zp_z)
\\
G_{{ \bf p}}(k_0)&=&
\frac{1}{{ \bf p}^2-k_0^2 -{\rm i} \Gamma^{\rm W}_{p_x}}
\end{eqnarray}
We can see from the diagonality of the Green function that only waves
which have a $k_x$ resonant with the structure will be transmitted into
the same $y,z$ mode of the waveguide surrounding the structure,
i.e. the
interferometer selectively transmits waves, it does not change their direction.

In the case where scatterers are present,
 there will be a self-energy $\Sigma^{\rm D}$ due to disorder
in addition to ${\rm i} \Gamma^{\rm pure}$.
One has
\begin{equation}
\label{gammapure+d}
 \Sigma_{{ \bf p}}=
\Sigma^{\rm pure}_{\bf p} + \Sigma^{\rm D}_{\bf p}
\approx
{\rm i}\Gamma^{\rm pure}_{{ \bf p}}+\Sigma^{\rm D}_{{ \bf p}}
\end{equation}
We are interested in the regime where the latter
has a much greater imaginary part,
 $\Gamma^{\rm D} $$\gg$$\Gamma^{\rm pure}$.
This is the case for a cavity with a very high intrinsic Q-factor
and a low density of scatterers.
Let us consider the transmission of a plane wave coming in from
$x=-\infty$, with wavevector ${ \bf a}$. 
(Note that in this section we consider transmission {\em through}
the cavity boundaries whereas in the previous section transmission
was {\em within} them, therefore the incoming nbeam is now in the 
$x$-direction.)
The momentum in the
$y$ and $z$ directions is quantized while
the momentum in the
$x$ direction outside the structure is determined by the total wavenumber,
$a_x=\sqrt{k^2-a_y^2-a_z^2}$.
\begin{equation}
\Psi_{\rm in}={\rm e}^{\textstyle {\rm i}{a}_{x} x}
\sqrt{\frac{4}{d_y d_z}} \sin({a}_{y} y) \sin({a}_{z} z)
\end{equation}

This incoming wave gives rise
to an intensity inside the structure
of
\begin{equation}
\label{dbb-incoming}
I({ \bf r})=\sum_{p_x} \frac{2p_x^2}{d_x} T^L_{{ \bf p}} |G_{{ \bf p}}|^2
|\Psi_{{ \bf p}}({ \bf r})|^2
\delta_{{p}_y,{a}_y}
\delta_{{p}_z,{a}_z}
\end{equation}
In the case no scatterers are present, the intensity of the transmitted
wave will be
\begin{equation}
I_{tr}=\sum_{p_x}\frac{p_x^2}{d_x^2} T^L_{{ \bf p}} T^R_{{ \bf p}}
|G_{{ \bf p}}|^2
\delta_{{p}_y,{a}_y}
\delta_{{p}_z,{a}_z}
\end{equation}
At resonance this transmission is unity for the case of symmetric,
non-absorbing barriers.
In practice, only one of the modes will contribute to this transmission.

In the case when scatterers are present, intensity will build up in the
cavity in essentially the same way, due to the condition $\ell >> d_x$,
but the dominant decay of the intensity will be due to scattering to other
 modes.
Hence, (\ref{dbb-incoming}) is still valid in this case, but we must take
the Green function with the disorder term $\Sigma^{\rm D}$ in the self-energy.
The incoming intensity will now be the source term for a BS equation which reads
\begin{equation}
\Phi_{{ \bf p}}=|G_{{ \bf p}}|^2 \left\{
\frac{2p_x^2}{d_x}T^L_{{ \bf p}}
\delta_{{p}_y,{a}_y}
\delta_{{p}_z,{a}_z}
+\sum_{{ \bf p'}}U_{{ \bf p}{ \bf p'}}\Phi_{{ \bf p'}}
\right\}
\end{equation}
$\Phi_{{ \bf p}}$ is the intensity in mode ${ \bf p}$, $U_{{ \bf p}{ \bf p}'}$ is
the irreducible vertex.
Since there are only discrete modes now, this is just a matrix inversion
problem:
\begin{equation}
\label{dbs-1d-matr}
\sum_{ \bf p'}
\left(
\delta_{{ \bf p}\,{ \bf p}'}
-
\frac{{\rm Im} \;G_{ \bf p}}
     {\Gamma_{ \bf p}}
U_{{ \bf p}{ \bf p'}}
\right)
\Phi_{{ \bf p}'}
=
|G_{{ \bf p}}|^2
\frac{2p_x^2}{d_x}T^L_{{ \bf p}}
\delta_{{p}_y,{a}_y}
\delta_{{p}_z,{a}_z}
\end{equation}
The matrix in the brackets has one eigenvalue close to zero which will
dominate the solution. To see this we multiply by $\Gamma_{ \bf p}$ and sum over
${ \bf p}$, using (\ref{Ward}) and (\ref{gammapure+d})
to find
\begin{equation}
\sum_{{ \bf p}}
 \Gamma^{\rm pure}_{{ \bf p}} \Phi_{{ \bf p}}
 =
 \sum_{{ \bf p}}
\Gamma_{ \bf p}
 |G_{ \bf p}|^2
\frac{2p_x^2}{d_x}T^L_{{ \bf p}}
\delta_{{p}_y,{a}_y}
\delta_{{p}_z,{a}_z}
\end{equation}
We expect the intensity in each mode to be proportional to the
density of states (this approximation is very good if the scattering rate
is much greater than the cavity loss rate). Also, if the incoming wave is
close to one resonant mode we can neglect contributions of the other modes.
This allows us to solve
(\ref{dbs-1d-matr})
and find
\begin{equation}
\Phi_{ \bf p}=
 \frac{\Gamma_{ \bf p} |G_{ \bf p}|^2}
  {\sum_{{ \bf p}'}
    \Gamma^{\rm pure}_{{ \bf p}'} \Gamma_{{ \bf p}'} |G_{{ \bf p}'}|^2}
\Gamma_{ \bf a}
 |G_{ \bf a}|^2
\frac{2{ \bf a}_x^2}{d_x}T^L_{{ \bf a}}
\end{equation}
The transmission from channel ${ \bf a}$ to ${ \bf b}$ is then given by
\begin{equation}
T_{{ \bf a}{ \bf b}}
=
\frac{\Phi_{ \bf b}}
 {2d_x}
T^R_{{ \bf b}}
\end{equation}
In case the double barrier structure is large ($d_x >> \ell$)
and uniform
in one or both of the
$(x,y)$ dimensions, the sum in the denominator can be simplified
since the resonance width $\Gamma_{p_x,p_y,p_z}$ will no longer
depend on the corresponding wavenumber.
For the case the structure is large in the $z$ dimension we find
\begin{equation}
\label{dbbtab}
T_{{ \bf a}{ \bf b}}=
\frac{
{ \bf a}_x^2
T^L_{{ \bf a}} \Gamma_{ \bf a} |G_{ \bf a}|^2
T^R_{{ \bf b}} \Gamma_{ \bf b} |G_{ \bf b}|^2
}
{
d_x^2
d_z
\sum_{p_x,p_y}
 \Gamma^{\rm pure}_{{ \bf p}}
 \nu_{{ \bf p}}
}
\end{equation}
with the definition of $\nu_{ \bf p}$ given in
(\ref{nu}). This is eq. (7) of ref. \cite{Theo}. 
In a more restricted version with one resonant mode
it was already derived in ref. \cite{BF};
note, however, that in that work the prefactor misses a factor 2.

In the case of a double barrier structure that is much larger than
 the bulk mean free path in two dimensions, ($d_x,d_y >> \ell$),
 the denominator of (\ref{dbbtab}) can be simplified to
 \begin{equation}
 4
 d_x^2
 d_y
 d_z
 \sum_{p_x}
 \Gamma^{\rm pure}_{p_x}
 (2+\frac{4}{\pi} \arctan \frac{k_0^2-p_x^2}{\Gamma_{p_x}})
\end{equation}

>From (\ref{dbbtab}), we see that scattering occurs into all free cavity
modes.  For a FPI $(d_y \to \infty)$ with incoming
plane wave, scattering inside the barrier
region causes an equal distribution over all directions. Therefore,
the transmitted intensity does not depend on the angle of incidence in the
$(y,z)$ plane. It does depend on the angle with the $x$ axis as
the mode quantization implies that only certain angles are transmitted.
This causes the well known appearance of fringes in the transmission
pattern. 

When a new cavity mode is just resonant, the return Green's function
in the cavity becomes large, as is discussed in section \ref{sec-subbot}.
This quenches scattering in the other subbands, so the device will
appear less disordered. This is one of the reasons why a
Fabry-Perot etalon gives best results when used near perpendicular incidence. 
We should point out, however, that this effect is less pronounced
for a two-dimensional structure such as a Fabry-P\'erot interferometer
than for a one-dimensional structure such as a wave guide,
as the divergence of the return Green's function is only logarithmic.

\subsection{Electronic conductivity through a disordered double 
barrier structure}
In \cite{Theo}
results were also derived for electronic double barrier quantum wells.
The conductance of such a device is again given by the
Landauer formula,
\begin{equation} 
G=\frac{2e^2}{h}\sum_{{ \bf a}{ \bf b}} {\cal T}_{{ \bf a}{ \bf b}}
\end{equation} 

In an experiment conducted by Gu\'eret et al, \cite{GRMM},
 the barriers  consist of two separated 
layers of
aluminium gallium arsenide (AlGaAs) in a gallium arsenide (GaAs) sample.

The thickness of the barriers
was varied from $7.5$ nm to $31$ nm. Hence, the transmission coefficients,
which depend exponentially on the barrier thickness, varied over
nine orders of magnitude, while the width of the resonance peak
remained a few  mV.
Elastic scattering from roughness at the interface
between the GaAs and the AlGaAs is believed to
cause this resonance broadening.

For a `pure' double barrier structure, the linewidth is expected to
scale linearly with the transmission coefficients:
\begin{equation} 
\Gamma=\Gamma^{\rm W}= 
\frac{k_0}{2d_x}( 
 A_{L}+ T_{L}
+ A_{R}+T_{R})
\end{equation} 
The total transmission for a resonant cavity is linear in the
transmission coefficients, as it is proportional 
to 
 $T_{\rm L}T_{\rm R} \Gamma^{\rm W}$. 

In the case of a small amount of disorder, for a cavity 
at resonance, the transmission coefficients can be seen from
(\ref{dbbtab}) to be proportional to 
 $T_{\rm L}T_{\rm R} \Gamma^{\rm W}$, as in the pure case.
The linewidth, however, is determined by $\Gamma^{\rm D}$,
which can be much larger.
It was calculated in \cite{Theo} that this multiple scattering
effect can explain the order of magnitude of the
observed resonance broadening.



\section{Strong scattering: the full Born series}
\markboth{Beyond second order Born}
         {Beyond second order Born}
\label{ch-beyondborn}

The results of section \ref{wguidemod} have been derived using the
second order Born approximation (\ref{tborn}) to the $t$-matrix.
This approximation essentially describes weak scatterers that have
no {\em internal} resonances near the wavelength of the incident waves.
Realistic systems, however, are not extremely long. If a reasonable amount of
scattering is to occur, one must have moderately strong scatterers.
But for such scatterers the multiple scatterings from the walls will
not at all be negligible. Rather than approximating the Born series
by the first two terms one has to sum the full series. For point scatterers
in continuum space this requires introducing either an extension of
wavefunction space \cite{TIP} or
a regularization of the return Greens function.
A very simple solution to this problem was given
in ref. \cite{NLT}. In the present section we shall apply
those ideas to  cavity systems. 

Scattering of Schr\"odinger waves in narrow systems has been studied 
by Bagwell \cite{BAGWELL}. He considers both finite size
 scatterers and point scatterers. In the latter case, the infinities 
that occur due to the small distance behavior are avoided by considering 
only a finite number of evanescent modes. As a result, the predictions of
the model then depend on the number of modes 
considered. Chu and Sorbello \cite{CHUSORB} 
investigated scattering properties of finite size scatterers
inside a cavity by means of the imaging method. 
We shall follow a different approach that is more suitable to
the limit of point scatterers, and then proceed to multiple 
scattering situations.

\subsection{Point scatterer near a mirror}

Boundary conditions can have a strong effect on the properties
of scattering, even for point scatterers. This can already be seen from
studying the following simple case: Consider a point scatterer for 
Schr\"odinger waves with `bare' scattering lenghth $u$
at a (small but finite) distance $d_w$ from  a perfect plane wall.
This imposes the mirror boundary condition $\Psi =0$ on the surface $z=0$.
The eigenfunctions of the pure system (without the scatterer) are 
\begin{equation}
\Psi(p_x, p_y, p_z) = \sqrt{2}\; {\rm i}\;
{\rm e}^{{\rm i}p_x x}
{\rm e}^{{\rm i}p_y y}
\sin (p_z z)
\end{equation}
with
\begin{equation} 
p_x^2+p_y^2+p_z^2=k_0^2
\end{equation} 
The return Green function is found by Fourier transformation:
\begin{equation}
\label{mirr-realsp-1}
G({ \bf r}_0,{ \bf r}_0)=G^0({ \bf r}_0,{ \bf r}_0)-G^0({ \bf r}_0,{ \bf r}_0^*)
\end{equation}
Where $\bbox{r}_0^{*}=(r^0_x,r^0_y,-r^0_z)$ is the mirror image of
${ \bf r}_0$, and $G^0$ is the free space 
propagator. 
We distinguish a `direct' and a `mirror' term. 
When the distance between the
scatterer and the mirror is much larger than the effective size
of the scatterer, $(d_w \gg \Lambda^{-1})$,
the mirror term has a regular behaviour.

We then need to regularize only the `direct' term in (\ref{mirr-realsp-1}),
the regularized `mirror' term will not be much different from the
unregularized form. This enables us to apply the regularization scheme
in \cite{NLT} to the direct term only, 
to find corrections due to the scattering between the 
impurity and its mirror image. It gives: 
\begin{equation}   
\label{returnfmirr}
G({ \bf r},{ \bf r})=\frac{{\rm i}k_0 +\Lambda}{4 \pi} 
-\frac{{\rm e}^{\textstyle 2 {\rm i}d_w k_0}}{8 \pi d_w}
\end{equation} 

However, when the distance between the mirror and the impurity is of
the order of the effective scatterer size,
this approach becomes ill behaved
and we need to regularize both terms in the return Green's function.  
We do this by applying a regulator of the form $\Lambda^2/(\Lambda^2+p^2)$
to the Fourier integrals that are
used to define the real space form of the Green's function.
This regulator is well known in quantum field theory. It cuts off momenta
beyond $p=\Lambda$, showing that $1/\Lambda$ is the effective size of the
scatterer. In the free space situation we thus get

\begin{eqnarray}
\label{grfcutoff}
G^{0}({ \bf r},{ \bf r'})&=&
\int \frac{{\rm d}^3{ \bf p}}{(2\pi)^3}
         \frac{1}{{ \bf p}^2-k_0^2-{\rm i}0}
\;
\frac{\Lambda^2}{\Lambda^2+{ \bf p}^2}
{\rm e}^{\textstyle {\rm i} { \bf p}\cdot ({ \bf r}-{ \bf r}')}\nonumber\\
&=&
\left\{
\frac{ {\rm e}^{\textstyle {\rm i}|{ \bf r}-{ \bf r}'|k_0}}
{ 4\pi |{ \bf r}-{ \bf r}'|}
-\frac{{\rm e}^{\textstyle -|{ \bf r}-{ \bf r}'|\Lambda}}
{ 4\pi |{ \bf r}-{ \bf r}'|}
\right\}
\frac{\Lambda^2}{\Lambda^2+k_0^2}
\end{eqnarray} 

Now we can take the limit ${ \bf r'} \to { \bf r}$. As expected, the result is finite,
and reads 
\begin{equation} 
\label{gregmetfactor}
G^0({ \bf r},{ \bf r})= \frac{{\rm i}k_0+\Lambda}{4\pi}
\;
\frac{\Lambda^2}{\Lambda^2+k_0^2}
\end{equation} 
The factor 
 $\Lambda^2/(\Lambda^2+k_0^2)$  is near unity for scatterers which are
much smaller than the wavelength. We will omit it from now on. Then
(\ref{gregmetfactor})  coincides with the result of ref \cite{NLT}:
\begin{equation} 
G^0({ \bf r},{ \bf r})= \frac{{\rm i}k_0+\Lambda}{4\pi}
\end{equation} 

The return Green's function (\ref{returnfmirr}) now reads
\begin{equation} 
G({ \bf r}_0,{ \bf r}_0)= \frac{{\rm i}k_0+\Lambda}{4\pi}
-\frac{{\rm e}^{ 2ik_0d_w}-{\rm e}^{ -2\Lambda d_w}}{8\pi d_w}
\end{equation} 
and reduces to eq. (\ref{returnfmirr}) when $\exp(-2\Lambda d_w)\ll 1$.
Now we use this regularized form of the return Green's function to
compute the $t$-matrix
\begin{equation}
\label{tmirror}
t=\left[ u^{-1}-\left(\frac{\Lambda+{\rm i}k_0}{4\pi} -
\frac{{\rm e}^{\textstyle 2{\rm i}d_w k_0}-{\rm e}^{\textstyle -2d_w\Lambda}}{8\pi d_w}
\right)
\right]^{-1}
\end{equation}
Following \cite{NLT} we state that by {\em definition}, a 
position dependent resonance occurs when
the $t$-matrix is purely imaginary. (This need not always coincide with
a sharp peak in the absolute value of the $t$ matrix). 
The positions of the resonances are located at $k_0$-values where
\begin{equation}
\label{respos}
\cos(2d_w k_0)=-8\pi d_w u^{-1}+2d_w \Lambda+\exp(-2d_w \Lambda)
\end{equation}
In case of Schr\"odinger waves, $u$ is the potential strength and
does not depend on $k_0$. This implies that
for large $d_w$ there is no resonance. Indeed, in free space $(d_w=\infty)$ it
is known that point scatterers for electrons do not have a resonance.
However, eq. (\ref{respos}) shows that an infinity of resonances occur
 when the scatterer is attractive ($u>0$) and is located close enough 
to the wall.


For scatterering of scalar classical waves, the analogue of the 
above position dependent resonance can be found by replacing the
scattering strength $u$  by
$4 \pi \alpha k_0^2$, where $\alpha$ is the  polarizability of
the point scatterer \cite{NLT}.
The expression for the $t$-matrix becomes
\begin{equation}
\label{tmirror-opt}
t=\left[ \frac{1}{4 \pi\alpha k_0^2}
-\left(\frac{\Lambda+{\rm i}k_0}{4\pi} +
\frac{{\rm e}^{\textstyle 2{\rm i}d_w k_0}-
{\rm e}^{\textstyle -2d_w \Lambda}}{8\pi d_w}
\right)
\right]^{-1}
\end{equation} 
It has resonances at
\begin{equation}
\label{respos-opt}
\cos(2d_w k_0)=
2 d_w
\left(
\frac{1}{\alpha k_0^2}
-\Lambda
\right)
+{\rm e}^{\textstyle -2d_w \Lambda}
\end{equation}
For $d_w \to \infty$ this yields the resonance condition
$k_0^2=k^2_*\equiv1/\Lambda\alpha$ of ref. \cite{NLT}.
For large but finite $d_w$ many resonances may occur.

\subsection{Pointscatterer in a Waveguide}
Next, consider a scatterer in an
infinitely long channel.
The Green's function for the clean channel is calculated in
(\ref{puregreen}).
It is expressed as a 
sum of mode propagators
\begin{eqnarray}
\label{channelprop}
G({ \bf r},{ \bf r}')= \sum_{{ \bf p}} \Psi_{{ \bf p}}(\bbox{\rho}) 
G_{{ \bf p}}(z-z') \Psi_{{ \bf p}}(\bbox{\rho}')
\end{eqnarray}  
The mode quantum numbers $p_x,p_y$ are the discrete transversal
wavenumbers.

For systems on a lattice the sum over ${ \bf p}$ converges as 
the distance between adjacent lattice points is a natural cutoff.

Even in a confined geometry of continuum space we have the problem that 
the real part of the
return Green's function diverges. In equation (\ref{channelprop})
this occurs since for large $|{ \bf p}|$ the terms in the infinite series
tend to zero as $1/p$, leading again to a linear divergency of the sum.
In order to define the $t$-matrix of the point scatterer in a meaningful way, 
we thus still need to regularize the return Green's function.

The approach followed by Bagwell \cite{BAGWELL} is to take into account only 
a finite number of evanescent modes in equation (\ref{channelprop}).
 As the cutoff has not been related to physical properties of the scatterer,
this approach is a bit unsatisfactory and unnatural.

 We can  regularize the sum (\ref{channelprop}) term by term 
in a way similar to (\ref{grfcutoff}), using the cutoff
function $\Lambda^2/(\Lambda^2+{ \bf p}^2+q^2)$, where ${ \bf p}$ is now 
the discrete transversal wavenumber and where
 $q$ is the continuous wavenumber in the $z$-direction.
\begin{eqnarray} 
\label{g-reg-wguide-1}
G_{{ \bf p}}(z,z')&=&\int_{-\infty}^{\infty} \frac{{\rm d}q}{2\pi}
\frac{1}{q^2+{ \bf p}^2-k_0^2-\Sigma_{{ \bf p}}}
\frac{\Lambda^2}{\Lambda^2+{ \bf p}^2+q^2}
{\rm e}^{{\rm i}q(z-z')}
\\ \label{g-reg-wguide-2}
&=&
\left(
\frac{{\rm i}{\rm e}^{\textstyle {\rm i}\sqrt{k_0^2-{ \bf p}^2+\Sigma_{{ \bf p}}}|z-z'|}}
{2 \sqrt{k_0^2-{ \bf p}^2+\Sigma_{{ \bf p}}}}
-
\frac{{\rm e}^{\textstyle -\sqrt{\Lambda^2+{ \bf p}^2}|z-z'|}}
{ 2 \sqrt{\Lambda^2+{ \bf p}^2}}
\right)
\frac{\Lambda^2}{\Lambda^2+k_0^2+\Sigma_{{ \bf p}}}
\end{eqnarray} 
We will consider the real space case again, omitting the factor
${\Lambda^2}/{(\Lambda^2+k_0^2+\Sigma_{{ \bf p}})}$ from now on.
The return Green's function will then be given by
\begin{eqnarray} 
\label{return-g-reg-wg}
G({ \bf r},{ \bf r})&=&
\sum_{{ \bf p}} \Psi_{{ \bf p}}^2(\bbox{\rho})
\left(
\frac{{\rm 1}}
{2 \sqrt{{ \bf p}^2-k_0^2-\Sigma_{{ \bf p}}}}
-
\frac{1}
{ 2 \sqrt{\Lambda^2+{ \bf p}^2}}
\right)
\end{eqnarray} 
For large $|{ \bf p}|$ the summand is of order $p^{-3}$ so
the series (\ref{channelprop}) converges.

For the special case where the energy $k_0^2$ approaches one of the subband
bottoms $p^2$ the series contains one divergent term. As we shall see 
in detail below, this divergency
will be regularized when the system is finite and there
is conductive behavior near the boundaries.

The $t$- matrix of the point scatterer 
(in a channel without further randomness)
 now reads
\begin{equation}
\label{tcleanchannel}
t=\left[
u^{-1} -
\sum_{{ \bf p}} \Psi_{{ \bf p}}^2(\bbox{\rho})
\left(
\frac{{\rm 1}}
{2 \sqrt{{ \bf p}^2-k_0^2-\Sigma_{{ \bf p}}}}
-
\frac{1}
{ 2 \sqrt{\Lambda^2+{ \bf p}^2}}
\right)
\right]^{-1}
\end{equation}
If the scatterer strength is positive, that is to say, for an {\em attractive}
scatterer potential, the $t$ matrix can have a pole.
It occurs before the onset of conduction modes, thus for $k_0< |{ \bf p}_0|$,
where ${ \bf p}_0$ is the label of the first mode. 
This bound state will have an energy between $0<k_0^2<{ \bf p}_0^2$,
 the energy of the first subband bottom, provided that 
\begin{equation}
\label{rescondition}
0<u<\left[\sum_{{ \bf p}} \Psi_{{ \bf p}}^2(\bbox{\rho})
\left(\frac{1}{2 \sqrt{{ \bf p}^2-\Sigma_{{ \bf p}}}}-
\frac{1}{ 2 \sqrt{\Lambda^2+{ \bf p}^2}}\right)\right]^{-1}
\end{equation}
Thus, the bound state already occurs for rather weak scatterers,
 which is an indication of the failure of low-order approximations, 
such as the second order Born approximation.
On the other hand, a {\em strong} attractive scatterer will have its bound
state at negative energy, $E=k_0^2<0$. In an optical situation such a
state has no meaning. In an electronic system it may occur, however. 
If the Fermi energy is above the bound state energy, the state will always be
occupied and it will not contribute to conductance. 
Due to the divergency of the return Green's function at the bottom of the
first subband, the bound state will always occur below the first subband
bottom. The weaker the scatterer is, the closer
the  bound state energy will move towards the first subband bottom. Also note
the explicit depencence on the transversal ($x,y$) position of the scatterer.
For very weak scatterers, however,
 the resonance will be indistinguishable from the
subband resonance and the second order Born approximation will yield 
good results.

\subsection{Resonant Tunneling}
Unexpected  peaks have been observed in the
transmission of quantum dots,  narrow constrictions
of a $2D$ electron gas \cite{McEuwen}. Before any of 
the channels of the dots were open,
a transmission peaks of order unity were observed. They were attributed to
tunneling through a bound state of a single impurity located (by accident) in
the channel. We model such a narrow junction between two broad reservoirs 
where conduction takes place by an infinite channel. We assume that the 
electrostatic potential equals $V_0$ for $0<z<L$ and vanishes 
elsewhere. This implies that for $z<0$ and for $z>L$ conduction occurs, 
while the `junction' $0<z<L$ acts as a barrier as long as
 $k_0^2<V_0$.

In the true $1D$ case this problem can easily be solved exactly,
yielding the amplitude transmission coefficient
\begin{equation} 
\label{mu-from-syst6}
\zeta=\frac{8{\rm i}k_0\kappa^2 {\rm e}^{ -\kappa L}}
{(2\kappa-u)(k_0+{\rm i}\kappa)^2-
uV_0[{\rm e}^{ -2\kappa Z_0}+{\rm e}^{ -2\kappa(L-Z_0)}]
+(2\kappa+u)(k_0-{\rm i}\kappa)^2{\rm e}^{ -2\kappa L}}
\end{equation} 

The intensity transmission coefficient is the absolute
square of the amplitude transmission coefficient $\zeta$. It reads,
 up to an exponentially small term in the denominator,
\begin{equation} 
\label{transmission-T}
 \displaystyle T= 
\displaystyle
\frac{64 k_0^2 \kappa^4{\rm e}^{ -2\kappa L}}
{4 k_0^2 \kappa^2(2 \kappa -u)^2 +[(2\kappa-u)(k_0^2-\kappa^2)
-uV_0({\rm e}^{ -2\kappa Z_0}+{\rm e}^{ -2\kappa(L-Z_0)})]^2}
\end{equation} 

Since the potential is equal on both sides of the barrier, and the incoming 
and outgoing modes are the same  the flux
transmission coefficient ${\cal T}$ equals the intensity transmission
coefficient $T$. 
When $k_0^2=k_*^2\equiv V_0-u^2/4$, there is a resonance peak in the 
transmission. Since then $\kappa=u/2$, its height is
\begin{equation} 
\label{1dtunnelt}
{\cal T}_{\rm max}={\cal T}(k_*)=\frac{2k_*^2u^2}
{V_0^2\left(1+\cosh u(L-2Z_0) \, \right)}
\end{equation} 
>From this equation we can see that the tunneling transmission is
maximal if the scatterer is positioned in the centre of the barrier,
$Z_0=L/2$. Then, the transmission coefficient is no longer exponentially
small. 
Independent of the
length of the barrier, its maximal value is
\begin{equation} 
{\cal T}_{\rm max}[L/2]=
\frac{k_0^2u^2}{V_0^2}=\frac{(V_0-u^2/4)u^2}{V_0^2}
=1-\left(\frac{V_0-u^2/2}{V_0}\right)^2
\end{equation} This phenomenon is called resonant tunneling.
We see that $T$ is unity if $u^2=2V_0$. This occurs when the resonance energy
is exactly half the barrier energy,
$k_*^2=V_0/2$. 
We will now show how this feature of a truly $1D$ system
 persists in the quasi- $1D$ waveguide.

\subsection{Resonant tunneling in a waveguide}

The situation for impurity tunneling through a barrier in a waveguide 
is analogous to the $1D$ case if we can assume that the
lowest subband is nondegenerate, i.e. there is a lowest subband
$p_0$ and all other subbands are located at a higher energy. In that
case, we can ignore tunneling through the higher subbands. We {\em do}
need to include all subbands in the calculation of the return 
Green's function, as well as reflection terms for the lowest subband.
>From (\ref{g-reg-wguide-2}), the return Green's function becomes
\begin{eqnarray}
G(\bbox{\rho}_0,Z_0,\bbox{\rho}_0,Z_0)&=&\sum_{{ \bf p}} \Psi_{{ \bf p}}^2(\bbox{\rho})
G_{{ \bf p}}(0) 
\\ \nonumber
&&
+ G_{{ \bf p}_0}(-Z_0) r_{ \bf p} G_{{ \bf p}_0}(Z_0)
+ G_{{ \bf p}_0}(L-Z_0)r_{ \bf p} G_{{ \bf p}_0}(-L+Z_0)
\end{eqnarray} 
We will use the abbreviations $\kappa_{ \bf p}$
for the imaginary wavenumber inside the structure,
$q_{ \bf p}$ for the wavenumber outside the structure,
$r_{ \bf p}$ for the internal (amplitude) reflection coefficient
at the edge of the barrier.
\begin{eqnarray} 
\kappa_{{ \bf p}}&\equiv&\sqrt{{ \bf p}^2+V_0-k_0^2}
\qquad
\kappa\equiv \kappa_{{ \bf p}_0}
\\ 
q_{{ \bf p}}&\equiv&\sqrt{k_0^2-{ \bf p}^2}
\qquad
q\equiv q_{{ \bf p}_0}
\\ 
r_{ \bf p}&\equiv&
-2\kappa_{{ \bf p}} 
\frac{q_{{ \bf p}}-{\rm i}\kappa_{{ \bf p}}}{q_{{ \bf p}}+{\rm i}\kappa_{{ \bf p}}}
\end{eqnarray}
Thus, we can calculate the return Green's function, including the
reflection terms
\begin{eqnarray}
 G(\bbox{\rho}_0,Z_0,\bbox{\rho}_0,Z_0)
&=&
\sum_{{ \bf p}} \Psi_{{ \bf p}}^2(\bbox{\rho})
\left(
\frac{1}
{2 \kappa_{{ \bf p}}}
-
\frac{1}
{ 2 \sqrt{\Lambda^2+{ \bf p}^2}}
\right)
\\ &&
-\Psi_{{ \bf p}_0}^2(\bbox{\rho})
\frac{{\rm e}^{\textstyle -2\kappa_{{ \bf p}_0}Z_0}
 +{\rm e}^{\textstyle -2\kappa_{{ \bf p}_0}(L-Z_0)}}
{2\kappa_{{ \bf p}_0}}
     \frac{q_{{ \bf p}}-{\rm i}\kappa_{{ \bf p}}}
    {q_{{ \bf p}}+{\rm i}\kappa_{{ \bf p}}}
\end{eqnarray} 
Neglecting the direct tunnelling term
we can calculate the transmission
coefficient for a barrier with a scatterer at position $(\bbox{\rho}_0,Z_0)$.
\begin{equation}
\label{Transmission-barrier}
T=\frac{16 q^2 \kappa^2 \tilde{u}^2 {\rm e}^{ -2\kappa L}}
{(2 \!  -  \! u\tilde{R})^2\kappa^2V_0^2
-\kappa \tilde{u} (2 \! - \! u\tilde{R}) V_0(q^2 \!- \! \kappa^2)
[{\rm e}^{ -2\kappa Z_0} \! + \! {\rm e}^{ -2\kappa(L-Z_0)}]
+\tilde{u}^2V_0^2[{\rm e}^{ -2\kappa Z_0} \! + \! {\rm e}^{ -2\kappa(L-Z_0)}]^2 }
\end{equation} 
with the abbreviations
\begin{eqnarray} 
\tilde{R} &\equiv& \sum_{{ \bf p}} \Psi_{{ \bf p}}^2(\bbox{\rho}_0)
\left(
\frac{1}
{2 \sqrt{{ \bf p}^2+V_0-k_0^2}}
-
\frac{1}
{ 2 \sqrt{\Lambda^2+{ \bf p}^2}}
\right)
\\
\tilde{u}&\equiv& u\Psi_{{ \bf p}_0}^2
\end{eqnarray} 
The resonance condition now is $u\tilde{R}=2$,
where the peak flux transmission coefficient becomes
\begin{eqnarray} 
{\cal T}_{\rm max}&=&\frac{8 q_{*}^2\kappa_{*}^2}
{V_0^2(1+\cosh 2\kappa_{*}(L-2Z_0)) }
\end{eqnarray} 
Again the maximum peak height occurs when the scatterer is located 
at the middle of the barrier
\[ 
{\cal T}_{\rm max}[L/2]=4q_{*}^2\kappa_{*}^2/V_0^2
=1-
\left(\frac{V_0-2k_{*}^2+2{ \bf p}^2}{V_0} \right)^2
\]
It depends on the transversal scatterer position and on the scatterer strength
only through the shift of the resonance position. Its 
maximal value is again unity.

To calculate the width of the resonance peak of
(\ref{Transmission-barrier}). To do this we will need
some derivatives:
\begin{eqnarray} 
\frac{\partial}{\partial k_0} \kappa_{{ \bf p}} &=&
-\frac{k_0}{\kappa_{{ \bf p}}}
\\
\frac{\partial}{\partial k_0} q_{{ \bf p}} &=&
\frac{k_0}{q_{{ \bf p}}}
\\
\frac{\partial}{\partial k_0} \tilde{R} &=&
k_0\sum_{{ \bf p}} \Psi_{{ \bf p}}^2(\bbox{\rho}_0) 
\frac{1}{2\kappa_{{ \bf p}}^3}
\approx 
 \Psi_{{ \bf p}_0}^2(\bbox{\rho}_0) \frac{k_0}{2\kappa_{{ \bf p}_0}^3}
\end{eqnarray} 
The approximation for the derivative of $\tilde{R}$ is good
if $\kappa_{{ \bf p}_0}$ is much smaller than the other
decay constants, which it is in a usual tunneling situation.

The denominator of (\ref{Transmission-barrier}) consists of three
terms, the first two are zero at resonance. 
Half height occurs approximately
when the sum of those terms equals the third term.
The first term of the denominator is of second order in 
\mbox{$\Delta k =k_0-k_{*}$}, the second term is of first order, but
has a much smaller coefficient, so we will have to include both terms
to 
 determine the peak width.
We solve
\begin{equation} 
\frac{(\Delta k)^2}{2} \tilde{u}^2\frac{k_0^2}{2\kappa^4} V_0^2 -
(\Delta k) \tilde{u}^2 \frac{k_0}{\kappa^2} (q^2-\kappa^2)
({\rm e}^{ -2\kappa Z_0} \! + \! {\rm e}^{ -2\kappa(L-Z_0)})
=\tilde{u}^2V_0^2({\rm e}^{ -2\kappa Z_0} \! + \! {\rm e}^{ -2\kappa(L-Z_0)})^2
\end{equation}  
to find the result
\begin{eqnarray} 
(\Delta k)_{\rm FWHM} &=& 
4 \kappa^2 
\frac{{\rm e}^{ -2\kappa Z_0} \! + \! {\rm e}^{ -2\kappa(L-Z_0)}}
{k_0V_0}\sqrt{q^4+\kappa^4}
\\
 &\approx& 
4 \kappa^2 
\frac{{\rm e}^{ -2\kappa Z_0} \! + \! {\rm e}^{ -2\kappa(L-Z_0)}}
{k_0}
\end{eqnarray} 
So the peak full width at half maximum, $(\Delta k)_{\rm FWHM}$,
is exponentially small even if the scatterer 
is positioned near the center.
Only through the
resonance energy, determined by $u\tilde{R}=2$, it  depends  on  the
$(x,y)$ coordinates of the scatterer or its scattering strength $u$.
 This means that of the three
quantities that are easily accessible by experiment, the resonance
position, heighth and width, only the resonance position contains
information about the scattering strength and $(x,y)$ position.

\subsection{Multiple scattering}
Suppose that in the region $0<z<d_z$ there is an a priori 
nonuniform density $n(x,y)$ of randomly placed, possibly resonant impurities.
 To first approximation, this causes the average 
subband Green's functions
to obtain a self-energy term
\begin{equation}
\label{sigmafirst}
\Sigma_{{ \bf p}}=\int {\rm d}^2 \bbox{\rho} 
|\Psi_{{ \bf p}}(\bbox{\rho})|^2 n(\bbox{\rho}) t(\bbox{\rho})
\end{equation}

The self-consistent $t$ matrix can now be calculated by
inserting the self-energy $\Sigma$ in (\ref{tcleanchannel}).
In order to describe diffusion, 
we need to define the ladder vertex $U$, which
must obey a local conservation law (Ward identity) of the form of 
(\ref{Ward})
\begin{equation}
\label{WWaarrdd}
{\rm Im}\Sigma_{{ \bf p}}=\sum_{{ \bf p}'}U_{{ \bf p}{ \bf p}'} 
{\rm Im} G_{{ \bf p}'}= 
\sum_{{ \bf p}'}U_{{ \bf p}{ \bf p}'}\nu_{ \bf p}
\end{equation}
This identity ensures flux conservation.
We define the ladder vertex $U$ as
\begin{eqnarray}
\label{fulltvertex}
U_{{ \bf p}{ \bf p}'}&\equiv& \int {\rm d}^2 \bbox{\rho} 
\Psi_{{ \bf p}}^2(\bbox{\rho}) \Psi_{{ \bf p}'}^2(\bbox{\rho}) n(\bbox{\rho}) 
t(\bbox{\rho}){t}^*(\bbox{\rho})
\end{eqnarray}
Then the Ward identity (\ref{WWaarrdd}) indeed is satisfied, since
\begin{equation} 
t({ \bf r})-\bar{t}({ \bf r})=\frac{1}{u^{-1}-G} -\frac{1}{u^{-1}-G^*}
=\frac{({u^{-1}-G^*})-({u^{-1}-G})}
{({u^{-1}-G})({u^{-1}-G^*})}
=tt^* (G-G^*)
\end{equation}
Where we have written $G$ for $G({ \bf r},{ \bf r})$.
With the vertex function defined in (\ref{fulltvertex}) we can
reproduce all results from section \ref{wguidemod}\ if we make the following
modifications:
\begin{itemize}
\item Wherever the expression \mbox{$k_0^2-{ \bf p}^2+i\Gamma_{{ \bf p}}$} is used,
      we need to replace it by \mbox{$k_0^2-{ \bf p}^2+\Sigma_{{ \bf p}}$},
      thereby incorporating the real part of the self-energy.
\item When the imaginary part $\Gamma_{{ \bf p}}$ is used outside the
      propagator denominator, we should replace it by 
      \mbox{${\rm Im \;} \Sigma_{{ \bf p}}$}.
\item The matrix $N_{{ \bf p}{ \bf p}'}$,
      the second order Born approximation to the
      irreducible vertex,  should be replaced by the full $t$-matrix
      approximation $U_{{ \bf p}{ \bf p}'}$ defined in (\ref{fulltvertex}). 
\end{itemize}

To find the correct self-energy we need to solve $t(\bbox{\rho})$ and
$G_{{ \bf p}}$ simultaneously, in general we will need to do this
numerically. As the equations are well-behaved in a physical situation,
an iterative
improvement method can be used. In section \ref{ch-sim}
 we will discuss an approach
for a lattice model where a direct simulation of the scattering
problem was performed.

The drops in the conductivity that can again be seen in the case
of attractive scatterer, shortly before the
opening of a new subband can be explained qualitatively as
single scatterer resonances. The real part of the Green's function
of the not-yet opened subband starts to diverge, which means
it has to cross a resonant value, rendering the $t$-matrices
of the individual scatterers imaginary and very large. Thus, the
scatterers scatter more eficiently at these energies, thereby
decreasing the conductance. As the resonance is position-dependent,
there will not be a sharp drop in conductivity, but instead a smooth
decrease.
When the new subband opens, the density of states effects will cause
a sharp drop in conductivity, which is already present in the second order
Born approximation.

For repulsive scatterers no resonance exists and the conductivity will 
rise with increasing energy until the opening of the nest subband, at which time
it will drop again.

\subsection{Subband bottom transparency}
\label{sec-subbot}
For the case of a single scatterer in a channel an interesting effect 
was observed  by Chu and Sorbello
\cite{CHUSORB} and by Bagwell \cite{BAGWELL}: when the energy is
 close to the bottom of a subband,
the return Green's function is very large so that the $t$-matrix is small. 
As a result, there is almost no scattering. 

For the multiple scattering situation the question that arises immediately
is: will the system become optically thin, so
that {\em ballistic} transport determines the transmission
coefficients?

Let the energy approach
a subband bottom $k_0^2 \approx { \bf p_*}^2$.
The return Green's function $G_{{\bf p}_*}$ of this mode will become large
due to its quasi $1-D$ square root divergency. 
The
corresponding term in the denominator of (\ref{tcleanchannel})
will dominate the other terms, except 
at the nodes of the corresponding subband wave function. 
If this wavefunction has no nodes, or the scatterer density
is zero near those nodes, as can be the case for a waveguide with disorder only
near the edges, the self energy
of the corresponding mode can be solved approximately:
\begin{eqnarray} 
\nonumber
\Sigma_{{ \bf p}_*}&=&\int {\rm d}^2 \bbox{\rho} 
|\Psi_{{ \bf p}_*}(\bbox{\rho})|^2 n(\bbox{\rho}) 
\frac{1}{u^{-1}-\sum_{{ \bf p}'}|\Psi_{{ \bf p}'}(\bbox{\rho})|^2G_{{ \bf p}'}(0)}
\\ \nonumber
&\approx&
\int {\rm d}^2 \bbox{\rho} 
|\Psi_{{ \bf p}_*}(\bbox{\rho})|^2 n(\bbox{\rho}) 
\frac{-1}{|\Psi_{{ \bf p}_*}(\bbox{\rho})|^2G_{{ \bf p}_*}(0)}
\\ \nonumber
&=&
\int {\rm d}^2 \bbox{\rho} 
 n(\bbox{\rho}) 
2{\rm i}\sqrt{k_0^2-{ \bf p}_*^2+\Sigma_{{\bf p}_*}}
\\
\label{sigmasubband}
&=&
2{\rm i} \, n_{\scriptscriptstyle 1D}\, \sqrt{k_0^2-{ \bf p}_*^2+\Sigma_{{\bf p}_*}}
\end{eqnarray} 
where $n_{\scriptscriptstyle 1D}$ is the $1D$ scatterer density
\begin{equation} 
n_{\scriptscriptstyle 1D}=\int {\rm d}^2 \bbox{\rho}\,\, n(\bbox{\rho})
\end{equation} 
We see that (\ref{sigmasubband}) does not contain the scatterer strength
$u$. It is valid for large values of $u$, as for
very small scattering strength the term $u^{-1}$ will not be
neglegible compared to the return Green's function.
Squaring (\ref{sigmasubband}) and requiring that the imaginary part of 
the self-energy stays positive, we find
\begin{equation} 
\Sigma_{{\bf p}_*}=-2\, n_{\scriptscriptstyle 1D}^2\, 
+2{\rm i}\,n_{\scriptscriptstyle 1D} \sqrt{k_0^2-{ \bf p}_*^2
-n_{\scriptscriptstyle 1D}^2}
\end{equation}
This expression is of second order in $n(\bbox{\rho})$.
Although our theory is
a first order approximation in $nt$,
 we can still use it because
 the $t$ matrix
is so small  near the subband bottom.
This is the multiple scattering analogue of the subband bottom
transparency effect noted by Chu and Sorbello
\cite{CHUSORB} and by Bagwell \cite{BAGWELL} for the case of a single
scatterer: when the energy is at the subband bottom there is very little
scattering, and the system may even become optically thin, in which
case {\em ballistic} transport determines the transmission
coefficients. 
We can calculate the behaviour of the mean free path, which is,
unlike the selfenergy, usually easily experimentally accessible. 
\begin{eqnarray}
\{\ell_{{\bf p}_*}\}^{-1} &=&
2 {\rm Im \,}
\sqrt{k_0^2-{ \bf p}_*^2
      -2\,n_{\scriptscriptstyle 1D}^2
      +2{\rm i}\, n_{\scriptscriptstyle 1D} \sqrt{k_0^2-{ \bf p}_*^2- 
n_{\scriptscriptstyle 1D}^2}}
\\
&=&
2 {\rm Im \,}\left({\rm i}\, n_{\scriptscriptstyle 1D}
                   +\sqrt{k_0^2-{ \bf p}_*^2- n_{\scriptscriptstyle 1D}^2} \right)
\end{eqnarray}
This gives immediately
\begin{eqnarray}
\ell_{*,\rm max}&=& \frac{1}{2\, n_{\scriptscriptstyle 1D}}
\end{eqnarray} 
If there are lower lying, already propagating modes, their
 mean free path can be much longer than this.
Let's first consider the case where the scatterer density $n(x,y)$ is
zero near the nodes of the mode wavefunction $\Psi_{{\bf p}_*}$.
Then we can find the self-energy of the other modes by
\begin{equation} 
\Sigma_{{ \bf p}}=
\Sigma_{{ \bf p}_*}
\int {\rm d}^2 \bbox{\rho} 
\;
\frac{n(\bbox{\rho})}{n_{\scriptscriptstyle 1D}}
 \frac{|\Psi_{{ \bf p}}|^2}
      {|\Psi_{{ \bf p}_*}|^2}
\end{equation} 
If we estimate the value of the integral to be roughly unity, 
 we
find for the mean free paths of the other modes
\begin{equation}
\label{approxothermodes}
\ell_{{ \bf p}}^{-1}
\approx
 2 {\rm Im \,}
 \sqrt{k_0^2-{\bf p}^2-2 n_{\scriptscriptstyle 1D}^2+
   2{\rm i}\, n_{\scriptscriptstyle 1D}\sqrt{k_0^2-{\bf p}_*^2-
 n_{\scriptscriptstyle 1D}^2}}
\end{equation} 
Since $k_0^2-{ \bf p}^2$ is much larger than the other terms we can
neglect terms of order  $(k_0^2-{ \bf p}_*^2)/(k_0^2-{ \bf p}^2)$ to
find 
\begin{equation} 
\{\ell_{{ \bf p}}\}^{-1}
\approx
2\,  n_{\scriptscriptstyle 1D} 
\frac{\sqrt{k_0^2-{ \bf p}_*^2- n_{\scriptscriptstyle 1D}^2}}
{\sqrt{k_0^2-{ \bf p}^2}}
\end{equation} 
Where  ${\bf p}_*^2 \approx k_0^2$ is the newly opened mode and ${\bf p}^2 <<k_0^2$
for the already open modes.
It is seen that this expression vanishes at the exact subband
bottom, so the mean free path diverges and the transmission becomes 
ballistic in every mode except the newly opened one.

In practice, the scatterer density will not in general be exactly zero
near the nodes of the wavefunctions. 
The $t$-matrix
may be appreciable  near the zeroes of the wavefunction $\Psi_{{ \bf p}_*}$ 
of the `new' mode, so that the approximation (\ref{approxothermodes})
cannot be made.
This will have non-neglegible effects on the
behaviour of the other modes that do not have the same zeroes.
The corrections to the self-energy of the propagating modes 
that arise will be  
 of order $1/\sqrt{G_{{ \bf p}_*}}$, 
which is very small near the subband bottom 
so the resulting mean free path may, 
with these corrections, still become much larger than in the middle of 
the band. If it approaches the order of the system length,
 we can no longer use the results derived for optically thick
systems. Thus, a sample may cross over from the semi-ballistic
regime of multiple scattering to the ballistic regime of `low-order' scattering
when the energy is near a subband bottom.

\section{Numerical Simulation}
\label{ch-sim}
In order to look numerically at the effect of scatterers in a wave guide
we use a model which contains the essential physics of the 
continuum Schrodinger equation whilst being ameanable to this 
kind of analysis.  
One such model is the Anderson Hamiltonian
on a square lattice.
\begin{eqnarray}
\nonumber
H&=&\sum_{n_x,n_y}
\left\{ -|n_x,n_y>\epsilon_{n_x,n_y}<n_x,n_y|\right. \\ 
\nonumber
         &+&\left. |n_x+1,n_y><n_x,n_y| + |n_x-1,n_y><n_x,n_y|  
         +|n_x,n_y+1><n_x,n_y| +|n_x,n_y-1><n_x,n_y| \right\}
\end{eqnarray}
By setting the hopping elements to unity we define the unit of energy. 
The model has an upper and lower 
limit of energy $E=[-4,4]$ for the ordered case $\epsilon_{n_x,n_y}=0$.
The electron like
band structure we are interested in starts at the band edge closest to $E=4$. 
The sign of the disorder
parameter $\epsilon_{n_x,n_y}$ has been  chosen such 
 that for positive values the scatterers are
attractive and negative values repulsive.  
The conductance of this system can be calculated 
by simply transforming the Hamiltonian above into a set of transfer 
matrices, one for each layer,
and multiplying them together to find the  
transfer matrix for 
the whole system.  However, in the strong scattering regime where
 the conductance is very 
small this method is unstable since half the eigenvalues grow 
exponentially and half decay. 
We therefore prefer to calculate the conductance from the Greens 
function which either decays or grows.
The method we use to calculate the Greens function
was introduced by Lee and Fisher \cite{leefisher} and given for 
the general case by MacKinnon \cite{mackinnon}.
It consists of using Dysons equation to calculate the required elements 
iterativley by adding sucessive disordered layers.  The general form for the
relation between  the Greens function elements and the
 transmission coefficients
has been given by Baranger and Stone \cite{barangerstone}
 and we use the Landauer-Buttiker formalism
to calculate the conductance from these coefficients.  

One of the major differences between the continuum model 
and the Anderson model
is in the number of evanescent modes in a complete basis at the 
Fermi energy.  The continuum model has an infinite number whereas the
 Anderson model only has
as many modes as there are lattice sites across the wire.
We can easily accomodate our continuum theory to this band structure by
 taking an appropriate 
momentum cutoff function.
In the simulations we present in this paper we have been interested in disordered
systems which are long in comparison to the mean free path but much narrower.
  In this
limit only the evanescent modes which have any appreciable probability 
amplitude 
between scattering events need be included.  This makes the Anderson model 
an ideal candidate 
for this calculation.  

Comparison between these numerical simulations and our continuum model can
 be made
without any free parameters since the scatterer size is fixed by the
 requirement
that each scatterer should occupy the volume of one lattice cell. 

In figure \ref{fig-1pct-attr} we compare the semi-ballistic theory
 (full line) with one
realization of the Anderson model (dotted line).
The sample configuration is that of a long, weakly disordered wire. 
In figure \ref{fig-1pct-repu} the same is shown for repulsive scatterers.
Both graphs also show the same curve calculated in the second order Born
approximation, in which there is no difference between attractive and repulsive
scatterers.  
The conductance is shown as a function of carrier energy, where the units
 of energy are
related to the band structure of the lattice ($-4=$ zero energy, $0$ band
 center).
It is seen that the subband maxima have a round shape for attractive
 scatterers
while they are quite sharp for repulsive scatterers. Also, the conductance 
is lower for 
attractive scatterers. This can be explained as an effect of the scattering
 resonances
that exist for attractive scatterers and are described in section
 \ref{ch-beyondborn}. 

The shape of the conductance graph is seen to be dependent on
the sign of the scattering strength,
its magnitude and the scatterer density. 
Depending on these parameters the
 conductance trace can be either smooth, 
sawtooth-like (fig \ref{fig-1pct-repu}),
or have steps (fig \ref{fig-bornfull}). These steps can be (much) smaller
than the universal height $2e^2/h$ of steps 
in the conductance of ballistic devices,
although they occur at the same energies.

It is also seen in these graphs that the second order Born approximation 
gives a conductance that is much lower than the full t-matrix, although
it takes into account less scattering diagrams. This effect is shown for
 two different 
scattering strengths
in fig. \ref{fig-bornfull}. The explanation is that contributions
of higher order scattering processes in the full t-matrix 
have the opposite sign of the lowest order contribution.

In figures \ref{fig-short-attr} and \ref{fig-short-rep} conductance of
shorter samples with a relatively high density of scatterers are shown.
The good agreement between our theory and the numerical date 
(average over 10 samples) demonstrates that our model, which assumes
pointlike scatterers  is still good
at finite volume fractions.

In figure {\ref{fig-rcontact}} we see the contact resistance term $R_c$ and
the extrapolation length $z_0$, which gives rise to this contact resistance. 
It is clearly seen that $z_0$ is highly nonuniform and varies within each
subband with a small dip attributed to scattering resonances (the return
propagator being small and real there), a peak owing to a weak form of
subband bottom
transparency (the propagator being large and real, so that scatterers cannot
be resonant and scattering is slightly suppressed)
 and a steep drop caused by the sudden increase in density of
states as the propagator crosses over from large and real to large and
 imaginary. Note that
the contact resistance $R_c$ has a much smoother behaviour. It occurs
since the conductivity $\sigma$ essentially varies in the same way as $z_0$

The extrapolation length $z_0$ is a weighted average of the mode mean
 free paths
(\ref{eq-z0})
so it is of the order of one mean free path. Since our theory presently
does not include quantum localization effects, it does not remain valid
for samples which are many  mean free paths in length. Therefore,
in the regime where the semi-ballistic theory
 is valid, the influence of the contact resistance
term in eq. (\ref{rplusrc}) will always be important.

Figure \ref{fig-toolong} shows an example of a wire which is so long
that localization effects can no longer be neglected

Nikoli\'c and MacKinnon \cite{NikMac} 
discuss the conduction of 
quantum wires with a large amount of boundary roughness
and a number of scattering `islands' in the bulk. 
They describe the wires with a tight binding lattice model, 
in which some lattice sites are nonconducting (infinite potential).
The conduction is then calculated numerically and averaged over
a large number of configurations.
Our model can be compared to their data for the case when there
is no boundary roughness. 
In their fig.8 our theory would predict a slightly higher
average conductance (about 1 times $ e^2/h$ higher) while the qualitative
shape of the curve is about the same. It should be noted that
a sample with these dimensions contains on the average only 5 (!) scatterers,
 and our extrapolation length $z_0$ is about $1.5$ 
times the sample
length $d_z$. Therefore this sample is only in the crossover region between 
the ballistic regime and the regime of semi-ballistic transport.

The data of Nikoli\'c and MacKinnon \cite{NikMac} 
also shows that the localization length
for samples containing these strong scatterers is 
only a few times larger than the mean free path,
which means that the range of validity of our 
analytic model, which presently does not contain
localization corrections, is rather limited.
Many experimentally accesible systems fall
within this range though.

Also, for weaker scatterers (with finite potentials)
the localization 
length will be longer, extending the range of validity
of our model to longer samples.

Although our model assumes the scatterers are in the bulk of the wire,
it can
also be used to study qualitatively the effects of surface roughness,
described as a high concentration of scatterers near the surfaces,
with a vanishing scatterer density in the bulk. 
In this case there will be a strong subband bottom 
transparency effect when the
second subband opens, as the scatterer density vanishes near the node
of the second subband wavefunction. This gives rise to a peak in the 
transmission of the wire at the exact bottom
of the second subband,
as can be clearly seen in figures 3 and 10
of \cite{NikMac}. 
This peak is now well explained by our theory.
 
\section{Discussion}

We have found expressions for average transmission properties of
two geometries of semi-ballistic devices, the waveguide-like
geometry and the double barrier structure. Both geometries
apply to optical systems and quantum-electronical systems.
Partly such systems are already experimentally
accessible, while other realizations are expected to become available
soon.

Steps in the conductance of electronic systems are predicted,
together with remarkable drops shortly before a new subband 
opens. These drops have been described before in ref (surke).
We notice that for systems containing pointlike scatterers, the behaviour
of the conductance near these drops depends on the sign of the
scattering length.

Our expressions go beyond the second order Born approximation
to the $t$-matrix of the individual scatterers (or random Gaussian
potentials) and can therefore
include scattering resonances. Divergencies are dealt with in a
physically meaningful way. It is shown that in a narrow
waveguide the second order Born approximation is not always good,
even when it is accurate for the same scatterers 
in wide threedimensional systems.

Comparing our theory to numerical solutions of an Anderson model
we find good agreement in the regime where localization effects
can be neglected.

\acknowledgments
The interest for this work arose from discussions with the late Shechao Feng,
which are gratefully acknowledged.
Th.M. N. also benefited from discussion with 
N. Garcia, J.J. Saens, B.A. Altshuler and I.V. Lerner, 
and for hospitality at the University of Madrid, where part of this work
was performed. 
His work was made possible by the Royal Dutch Academy of Arts and
Sciences (KNAW). This work is also supported by the NATO, grant CRG 921399.

\clearpage

\begin{figure}
\caption{The waveguide (with cross secion $d_x d_y$)
is infinitely long but the disorder is
only present in a finite section of length $d_z$. The infinitely 
long `pure'  sections act as ideal quantum leads.}
\label{fig-wguide}
\end{figure}

\begin{figure}
\caption{Conductivities of semi-ballistic devices with very low disorder, in the
second order Born approximation.
 The solid line
represents the conductivity of a disordered 
film of width $d$ as it goes through
multiples of the resonant width $d^*$. The dashed line gives the
conductivity of a $1$D constriction of a $2$D electron gas as 
a function of the width $d$. }
\label{Theofig}
\end{figure}

\begin{figure}
\caption{
A double barrier quantum well, or its optical analogon,
 the Fabri Perot Interferometer,
is modeled by a pair of delta function potential barriers 
in the $x=0$ and $x=d_x$ planes.
Between these `mirrors' impurities are present. 
Transport occurs in the $x$ direction, tunneling through the barriers}
\label{fig-dbqw}
\end{figure}

\begin{figure}
\caption{
Conductance of a long quantum wire containing 
a small density of strong attractive scatterers
as a function of carrier energy $E/V$.
The vertical lines indicate positions of the subband edges
in the band structure of the lattice. 
}
\label{fig-1pct-attr}
\end{figure}

\begin{figure}
\caption{
Conductance of a long quantum wire containing 
a small density of strong repulsive scatterers
as a function of carrier energy.
The vertical lines indicate positions of the subband edges
in the band structure of the lattice. 
}
\label{fig-1pct-repu}
\end{figure}

\begin{figure}
\caption{
Conductance of a long quantum wire with a high density of 
strong scatterers.
At this point our simple model breaks down as localization
corrections become important. The effective mean free path is in the
order of one tenth of the wire length.
}
\label{fig-toolong}
\end{figure}

\begin{figure}
\caption{Extrapolation length z0 and 
contact resistance $R_{\rm c}=z_0/\sigma$
of a modestly disordered wire.}
\label{fig-rcontact}
\end{figure}

\begin{figure}
\caption{Conductance of short quantum wires with a large amount of attractive scatterers.
The thick line represents the semi-ballistic theory, the thin line is an average over
10 numerical simulations.
}
\label{fig-short-attr}
\end{figure}

\begin{figure}
\caption{
Conductance of short quantum wires with a large amount of repulsive scatterers.
The thick line represents the semi-ballistic theory, the thin line is an average over
10 numerical simulations.
}
\label{fig-short-rep}
\end{figure}

\begin{figure}
\caption{Conductance of a 1000x11 quantum wire with 5\% attractive scatterers,
 calculated using the full t-matrix (dashed curves) or the second order
Born approximation(dotted curves). For a bare scattering strength of 0.03
the second order Born approximation is a rather good approximation to the full
t-matrix, for u0=0.2 there is a large difference.}
\label{fig-bornfull}
\end{figure}

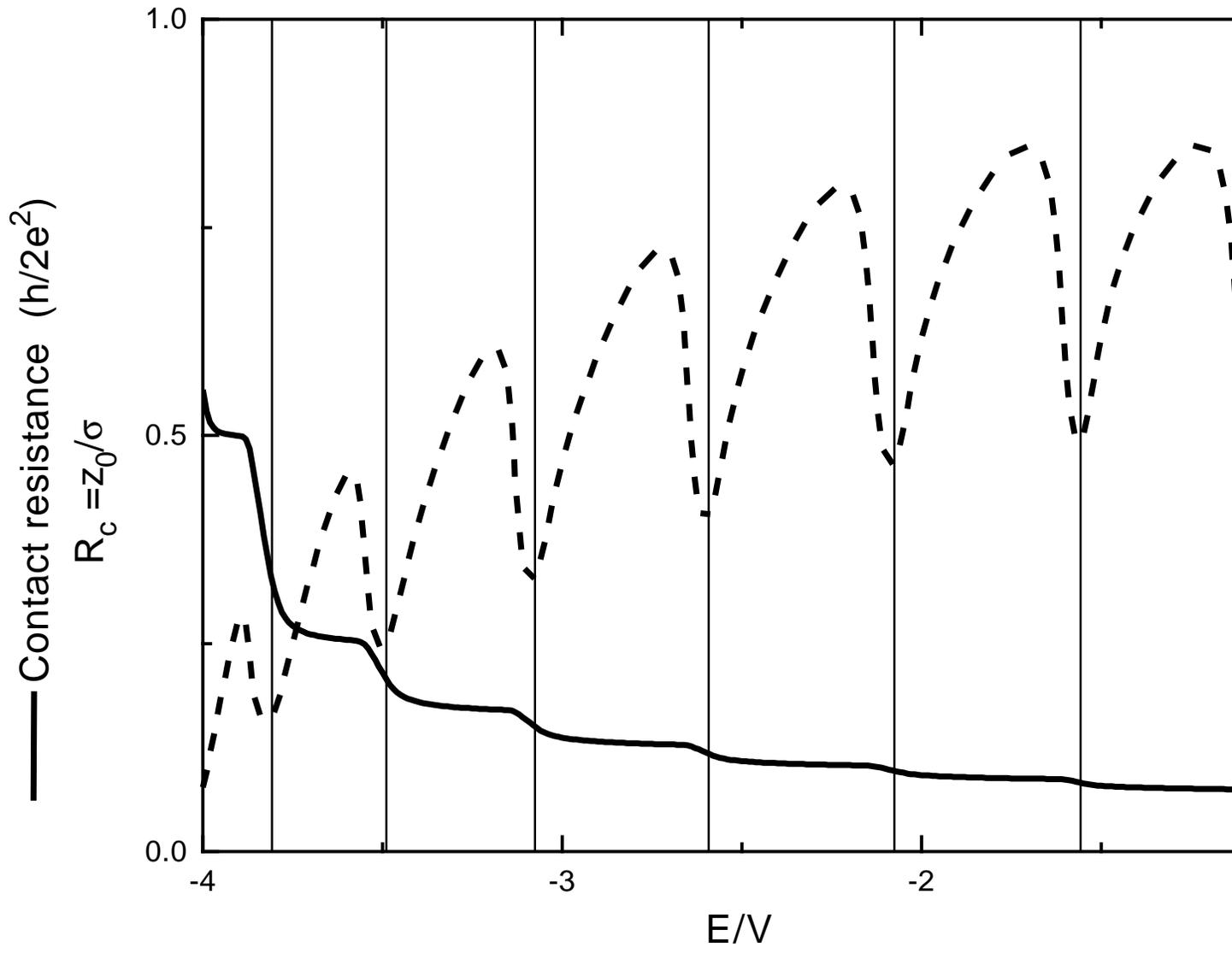
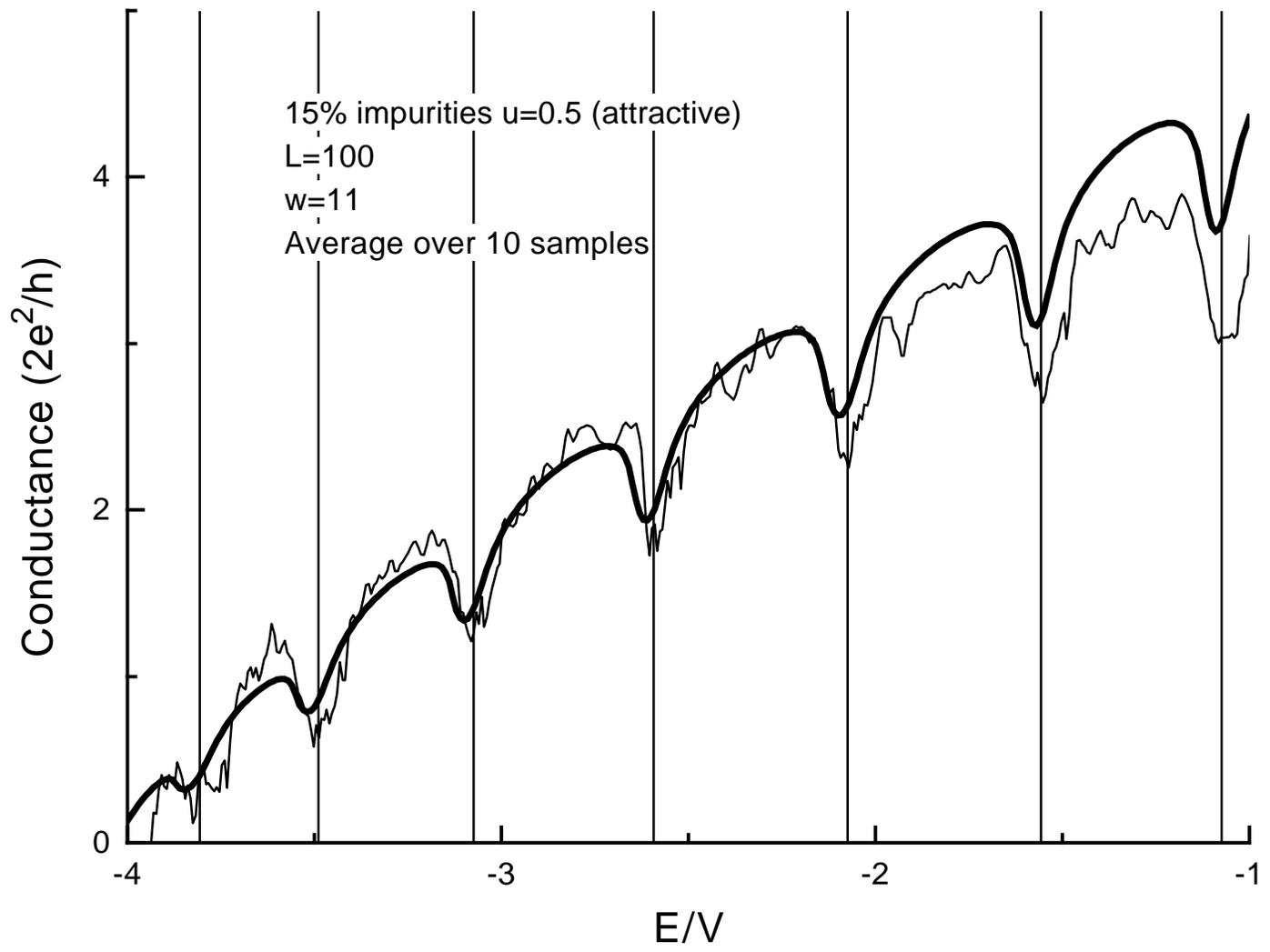
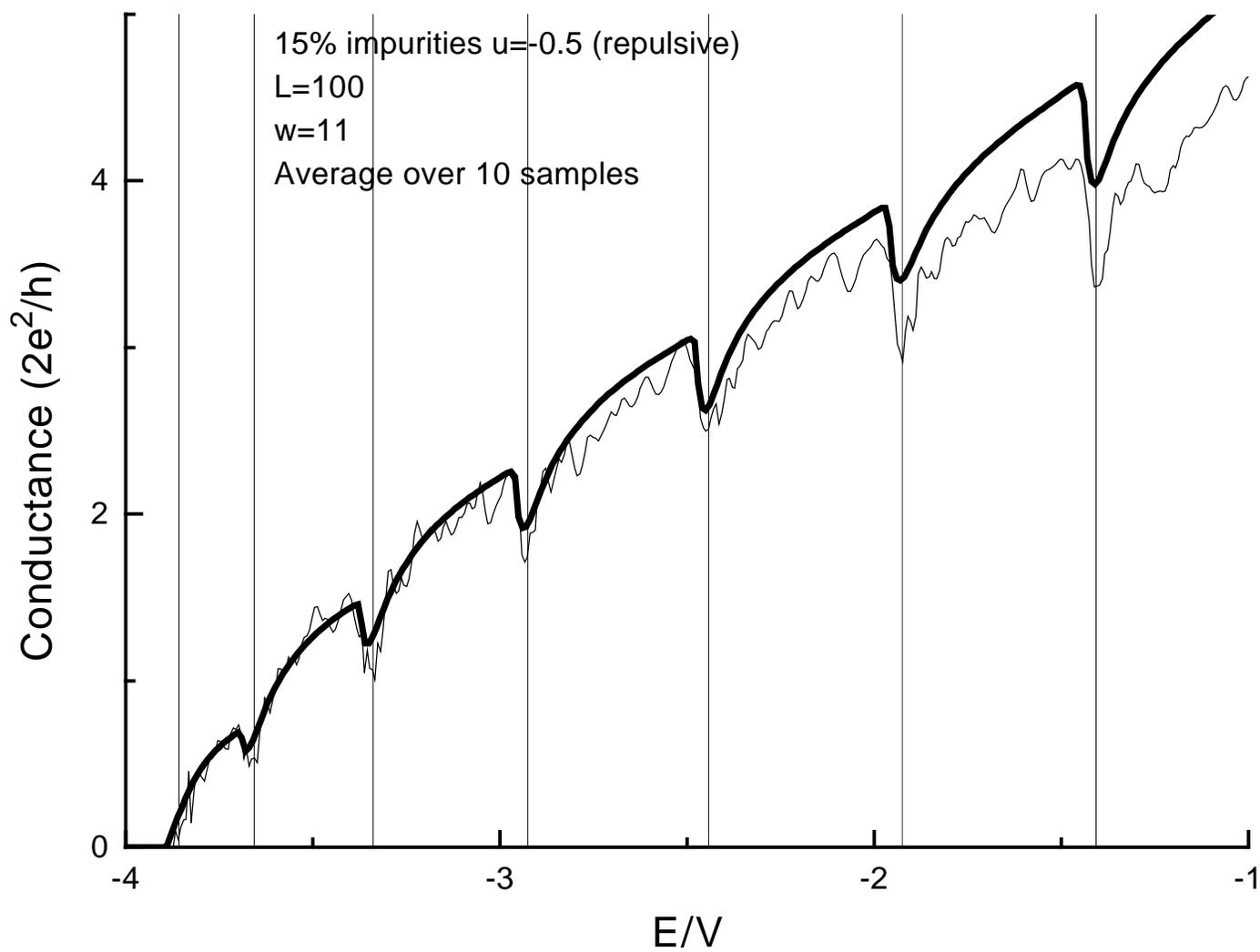
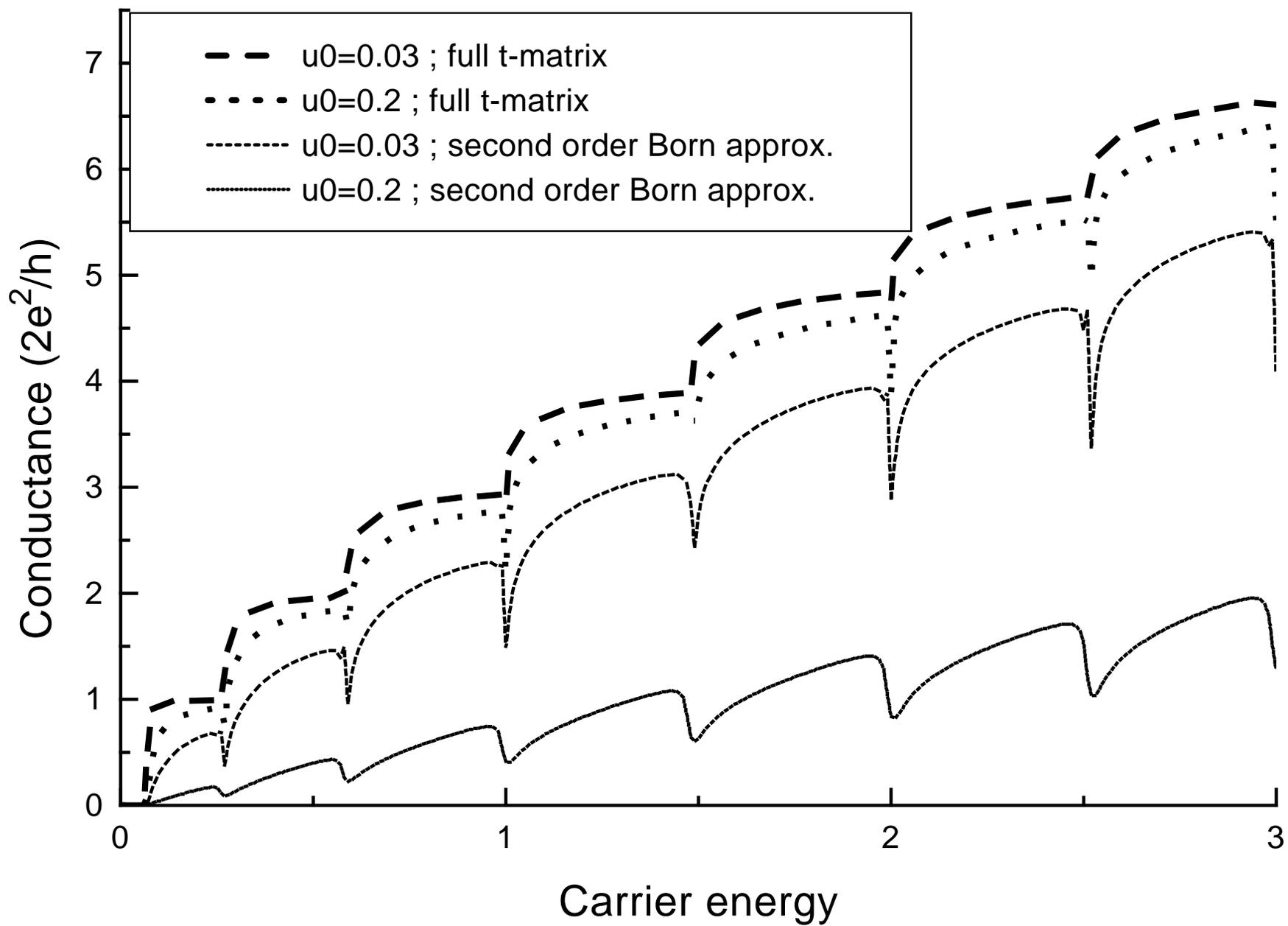
\end{document}